\documentclass[aps,prl,twocolumn,nofootnoteinbib,superscriptaddress]{revtex4-2}

\usepackage{epsf,amsmath,amssymb,verbatim,color,multirow,pifont,graphicx,cleveref,tabularx,cancel,soul,enumitem,arydshln,float}
\usepackage[dvipsnames]{xcolor}

\begin{document}
\title{Equilibrium-preserving Laplacian renormalization group}
\author{Sudo  \surname{Yi}}
\affiliation{School of Computational Sciences, Korea Institute for Advanced Study, Seoul 02455, Korea}
\affiliation{CCSS, KI for Grid Modernization, Korea Institute of Energy Technology, 21 Kentech, Naju, Jeonnam 58330, Korea}
\author{Seong-Gyu  \surname{Yang}}
\affiliation{School of Computational Sciences, Korea Institute for Advanced Study, Seoul 02455, Korea}
\affiliation{Integrated Science Lab, Department of Physics, Ume\r{a} University, SE-901 87, Sweden}
\author{K.-I. \surname{Goh}}
\email{kgoh@korea.ac.kr}
\affiliation{Department of Physics, Korea University, Seoul 02841, Korea}
\author{D.-S. \surname{Lee}}
\email{deoksunlee@kias.re.kr}
\affiliation{School of Computational Sciences, Korea Institute for Advanced Study, Seoul 02455, Korea}
\affiliation{Center for AI and Natural Sciences, Korea Institute for Advanced Study, Seoul 02455, Korea}

\begin{abstract}
Diffusion over networks has recently been used to define spatiotemporal scales and extend Kadanoff block spins of Euclidean space to supernodes of networks in the Laplacian renormalization group (LRG). Yet, its \textit{ad hoc} coarse-graining procedure remains underdeveloped and unvalidated, limiting its broader applicability. Here we rigorously formulate an LRG preserving the equilibrium state, offering a principled coarse-graining procedure. We construct the renormalized Laplacian matrix preserving dominant spectral properties using a proper, quasi-complete basis transformation and the renormalized adjacency matrix preserving mean connectivity from equilibrium-state flows among supernodes. Applying recursively this equilibrium-preserving LRG to various hypergraphs,  we find that in hypertrees with low spectral dimensions vertex degree and hyperedge cardinality distributions flow toward Poissonian forms, while in hypergraphs lacking a finite spectral dimension they broaden toward power-law forms when starting from Poissonian ones, revealing how informational, structural, and dynamical scale-invariances are interrelated.
\end{abstract}
\date{\today}

\maketitle 

{\it Introduction.---}Networks~\cite{Barabasibook} derived from empirical data are  obtained under finite resolution constraints and thus often incomplete. Inferring missing information can help mitigate this incompleteness~\cite{Zhou2009}. An alternative approach is to investigate how structural features evolve under varying resolution to infer them at resolutions other than currently available. Generating low-resolution networks, as in the renormalization group (RG)~\cite{Cardy1996}, can facilitate structural comparisons across scales. Similar to Kadanoff block spins in Euclidean lattices, networks can be coarse-grained by merging nodes within a given network distance~\cite{Song2005, PhysRevE.72.045105, Goh2006} or geographically proximate in an embedding space~\cite{Kim04,Serrano2008} into blocks. Such spatial proximity of nodes is, however, not always well-defined for real-world networks given their small-world nature~\cite{Watts:1998qf} and heterogeneous connectivity~\cite{Barabasi1999_BA}.

Among efforts to overcome these limitations~\cite{Arenas_2007,Gfeller2007,PhysRevE.89.052815,Garuccio2023,Villegas:2023aa,thibeault_low-rank_2024,poggialini2024,Gabrielli2024,Cheng2024,Nurisso2025,Jung2024}, spectral proximity provides a compelling alternative. Merging nodes with similar contributions to slow eigenmodes has been shown to allow the renormalized random walk generator to preserve long-timescale dynamics~\cite{Gfeller2007}. More recently, the Laplacian renormalization group (LRG)~\cite{Villegas:2023aa,poggialini2024,Gabrielli2024,Cheng2024,Nurisso2025} was introduced to obtain effective network representations for diffusion dynamics by merging nodes with sufficient diffusion between them into a {\it supernode}.  Implementing coarse-graining by utilizing the increasing irrelevance of fast diffusion modes over time, the LRG enables network exploration across spatiotemporal scales. However, the coarse-graining procedure remains \textit{ad hoc} and its formal procedure and validation remain incomplete, limiting its broader applicability. Most critically, as in the equilibrium RG where the partition function is preserved, the renormalized diffusion equation must accurately reproduce long-time  behavior, ensuring the invariance of the equilibrium state~\cite{Gfeller2007}. Preserving only the small eigenvalues of the Laplacian  would not be sufficient; the correct RG transformation of Laplacian using a proper basis---a quasi-complete one---is essential for its validity. Here, we present a rigorous formulation how to construct such a basis, offering the principled way of performing proper RG transformation of both Laplacian and adjacency matrices and coarse-graining of the system.

Given that RG transformations generally involve renormalizing both elements and their interactions, hypergraphs~\cite{Battiston:2020aa,Jhun:2019aa,Mulas:2022aa,Cooper:2013aa,Carletti:2021aa}---incorporating hyperedges that connect an arbitrary number of vertices---provide a natural mathematical framework for representing autonomously such transformations. Applying our equilibrium-preserving version of the  LRG to various hypergraphs with diffusion processes across vertices and hyperedges~\cite{Zhou2007,Chitra2019,Carletti2020,Nurisso2025}, we observe distinct structural evolution depending on spectral dimensionality: Degree and cardinality distributions converge to Poissonian forms in hypergraphs with low spectral dimensions, while they broaden toward power-law forms when starting from Poisson distributions in those lacking a well-defined spectral dimension, highlighting their fundamental difference at large spatiotemporal scales.
 
{\it Supernodes representing slow modes of diffusion.---}Let us consider a diffusion process on a network
\begin{equation}
    \frac{\partial x_i}{\partial t} = \sum_{j\in \mathcal{N}} A_{ij}x_j - \sum_{j\in \mathcal{N}}A_{ji} x_i=-\sum_{j\in \mathcal{N}} L_{ij}x_j,
    \label{eq:diff_def}
\end{equation}
in which $x_i$ is a dynamic variable at node $i\in \mathcal{N}$, where $N=|\mathcal{N}|$ is the number of nodes, with $\mathcal{N}$ being the set of nodes. The Laplacian is defined as $L_{ij} = q_i \delta_{ij} - A_{ij}$, where $A$  is the adjacency matrix  and $q_i\equiv\sum_{j\in \mathcal{N}}A_{ji}$ denotes the connectivity of node $i$. Employing the bra-ket notation, Eq.~\eqref{eq:diff_def} can be considered as a representation of an operator equation $\frac{\partial}{\partial t} |x\rangle = - L |x\rangle$ in the basis $\{|i\rangle | i\in \mathcal{N} \}$ and its dual $\{\langle i|\}$ with $x_i = \langle i|x\rangle$ and $L_{ij} = \langle i|L|j\rangle$. Its solution is $|x(t)\rangle = \rho(t) |x(0)\rangle$ with $|x(0)\rangle$ the initial state and $\rho(t)\equiv e^{-Lt}$ the (unnormalized) diffusion propagator. 

As $t$ increases, only slow eigenmodes, associated with small eigenvalues, of the Laplacian can contribute to the propagator as $\rho(t) \approx \sum_{\lambda\lesssim t^{-1}} e^{-\lambda t} |\lambda\rangle \langle \lambda|$ with $|\lambda\rangle (\langle \lambda|)$ denoting the right (left) eigenvector for an eigenvalue $\lambda$. One may expect that such slow relevant eigenmodes can be effectively represented by a small number of supernodes that form a partition of the original nodes. Consider a set of supernodes $\mathcal{N}'=\{i'\}$ with each original node $i$ belonging to exactly one supernode $i'$ ($i\in i'$). 
These supernodes can approximate well late-time diffusion dynamics if their associated orthonormal basis $\{|i'\rangle|i'\in \mathcal{N}'\}$, (i.e., $\langle i'|j'\rangle=\delta_{i'j'}$)  approximately spans the subspace of the slow diffusion modes. In other words, if the eigenvectors corresponding to small eigenvalues $\lambda < t_c^{-1}$ for given $t_c$ are well represented as 
\begin{equation}
    |\lambda \rangle \approx \sum_{i'\in \mathcal{N}'} |i'\rangle\langle i'|\lambda\rangle \ {\rm and} \ 
    \langle \lambda | \approx \sum_{i'\in \mathcal{N}'}\langle\lambda|i'\rangle \langle i'|,
    \label{eq:qc}
\end{equation}
then the propagator can be approximated by $\rho(t \gg t_c)\approx  \sum_{i',j'\in \mathcal{N}'} \rho_{i'j'}(t)|i'\rangle \langle j'| $ with  $\rho_{i'j'}(t) \equiv \sum_\lambda e^{-\lambda t}\langle i'|\lambda\rangle \langle \lambda|j'\rangle$ representing the renormalized propagator in the supernode basis. This {\it quasi-completeness} property  in Eq.~\eqref{eq:qc} is a fundamental requirement for the validity of the real-space LRG~\cite{Villegas:2023aa}, ensuring a reduced set of effective nodes to reproduce long-time diffusion dynamics.  

The physical meaning of supernodes is disclosed by Eq.~\eqref{eq:qc}. For any two nodes  $i$ and $j$ belonging to the same supernode $i'$,  the ratios of their propagator elements $\frac{\rho_{ik}(t)}{\rho_{jk}(t)}\approx \frac{\langle i|i'\rangle}{\langle j|i'\rangle}$ and $ \frac{\rho_{ki}(t)}{\rho_{kj}(t)}\approx \frac{\langle i'|i\rangle}{\langle j'|j\rangle}$ are independent of $k$ or $t$ when $t\gg t_c$. See Supplemental Material (SM)~\cite{SM} for derivation. This implies that nodes $i$ and $j$ are in equilibrium with each other for $t\gg t_c$, which we call {\it local equilibrium}, as analogy to the global equilibrium state where the propagator ratios for any two nodes are constant as  
$\lim_{t\to\infty} \frac{\rho_{ik}(t)}{\rho_{jk}(t)} = \frac{\langle i|0\rangle}{\langle j|0\rangle}$ and $\lim_{t\to\infty} \frac{\rho_{ki}(t)}{\rho_{kj}(t)} =\frac{\langle 0|i\rangle}{\langle 0|j\rangle}$ with $|0\rangle$ the eigenvector for $\lambda=0$ of the Laplacian. Note also that $\frac{\rho_{ij}(t) \rho_{ji}(t)}{\rho_{ii}(t)\rho_{jj}(t)} \approx 1$ for $i$ and $j$ being in local equilibrium, which is similar to the condition used in Ref.~\cite{Villegas:2023aa} and will be employed here for constructing supernodes.

{\it Renormalized diffusion equation.---}Applying $\langle i'|$ to the diffusion equation in operator form and using Eq.~\eqref{eq:qc}, we obtain a renormalized  diffusion equation~\cite{SM}
\begin{equation}
    \frac{\partial x'_{i'}}{\partial t^\prime} = -\sum_{j'} L'_{i'j'}x'_{j'},
    \label{eq:diff_rg}
\end{equation}
generated by the renormalized Laplacian $L'_{i'j'} \equiv \sum_{i\in i', j\in j'} \langle i'|i\rangle L_{ij} \langle j|j'\rangle$ with time rescaled to match supernode-scale dynamics. The eigenvalues $\lambda'$ and eigenvectors $|\lambda'\rangle$ and $\langle \lambda'|$ of $L'$  approximate the small eigenvalues and corresponding eigenvectors of $L$ from Eq.~\eqref{eq:qc}. The basis transformation coefficients $\langle i'|i\rangle$ and $\langle j|j'\rangle$ derived below are key to ensuring this approximation.

Recall that the propagator ratio $\frac{\rho_{ik}(t)}{\rho_{jk}(t)}$ for nodes $i$ and $j$ remains constant during their local equilibrium, equal to $\frac{\langle i|i'\rangle}{\langle j|i'\rangle}$ with $i'$ the supernode containing both $i$ and $j$. In the global equilibrium state, the ratio is fixed at $\frac{\langle i|0\rangle}{\langle j|0\rangle}$.  Since these  two constants must be identical, it follows that $\langle i|i'\rangle \propto \langle i|0\rangle$,  and similarly $\langle i'|i\rangle \propto \langle 0|i\rangle$.  Therefore the supernode basis is related to the original basis  by
\begin{equation}
    |i'\rangle =  \sum_{i\in i'} |i\rangle \frac{\langle i|0\rangle}{\langle i'|0\rangle} \ {\rm and} \ 
    \langle i'| = \sum_{i\in i'} \frac{\langle 0|i\rangle}{\langle 0|i'\rangle}\langle i|,
    \label{eq:iprime}
\end{equation}
where the constants $\langle i'|0\rangle$ and $\langle 0|i'\rangle$ represent the projections of $|0\rangle$ and $\langle 0|$ onto the supernode $i'$, satisfying
$\langle i'|0\rangle \langle 0|i'\rangle  = \sum_{i\in i'}\langle i|0\rangle \langle 0|i\rangle$ so that $\langle i'|i'\rangle=1$. Using Eq.~\eqref{eq:iprime}, we find
\begin{equation}
    L'_{i'j'}     
    = \sum_{i\in i', j\in j'} \frac{\langle 0|i\rangle}{\langle 0|i'\rangle}  L_{ij}\frac{\langle j|0\rangle}{\langle j'|0\rangle}.
    \label{eq:Lprime}
\end{equation} 
Then the equilibrium state is preserved, i.e., $|0'\rangle =|0\rangle$  since $\sum_{j'} L'_{i'j'}\langle j'|0\rangle = \sum_{i\in i',j\in \mathcal{N}} \frac{\langle 0|i\rangle}{\langle 0|i'\rangle} L_{ij} \langle j |0\rangle=0$. Similarly, the left eigenvector is invariant, i.e., $\langle 0'| =\langle 0|$  since $\sum_{i'} \langle 0|i'\rangle L'_{i'j'}= \sum_{i\in \mathcal{N},j\in j'} \langle 0|i\rangle  L_{ij} \frac{\langle j|0\rangle}{\langle j'|0\rangle}  =0$.

The second smallest eigenvalue $\lambda_2'$ may differ from that of the original Laplacian since the supernode basis is only quasi-complete. As $\lambda_2$ and $\lambda_2'$ characterize the relaxation time, we rescale time as
\begin{equation}
t' = {\lambda_2 \over \lambda_2'} t
\label{eq:tprime}
\end{equation}
to match the relaxation time between Eqs.~\eqref{eq:diff_def} and \eqref{eq:diff_rg}. Therefore we expect $x'_{i'}(t')$ from Eq.~\eqref{eq:diff_rg} to approximate the coarse-grained original solution $x_{i'}(t) = \langle i'|x(t)\rangle=\sum_{i\in i'} \langle i'|i\rangle x_i(t)$ with $\langle i'|i\rangle=\frac{\langle 0|i\rangle}{\langle 0|i'\rangle}$ from Eq.~\eqref{eq:iprime} in the late-time regime. 

{\it Coarse-grained networks.---}Since  $L'_{i'i'} \neq \sum_{j'\neq i'} L'_{j'i'}$ (or equivalently $\sum_{j'} L'_{j'i'}\neq 0$), Eq.~\eqref{eq:diff_rg} cannot be readily decomposed into link flows as in Eq.~\eqref{eq:diff_def},  raising a question about the renormalized adjacency matrix $A'$. In equilibrium, where $x'_{i'} \propto \langle i'|0\rangle$,  Eq.~\eqref{eq:diff_rg} can be represented as $\frac{\partial x'_{i'}}{\partial t'} = \langle 0|i'\rangle^{-1} \sum_{j'} (f'_{i'j'}- f'_{j'i'})=0$ in terms of rescaled flows  $f'_{i'j'} \equiv - \langle 0|i'\rangle L'_{i'j'}\langle j'|0\rangle$ satisfying $\sum_{j'} f'_{j'i'} = 0$ from Eq.~\eqref{eq:Lprime}. Moreover, the rescaled flows are aggregated through coarse-graining as $f'_{i'j'}=\sum_{i\in i', j\in j'} f_{ij}$ for $i'\neq j'$ with $f_{ij}\equiv -\langle 0|i\rangle L_{ij} \langle j|0\rangle$, suggesting that  these flows can effectively characterize the structure of the coarse-grained networks.

Assume that the original Laplacian is represented as $L_{ij} = \frac{\delta_{ij} q_i - A_{ij}}{N\langle 0|i\rangle \langle j|0\rangle}$.
Then the adjacency matrix is represented as $A_{ij}=N f_{ij} = -N \langle 0|i\rangle \langle j|0\rangle L_{ij}$ for $i\neq j$. Notice that this assumption holds for Eq.~\eqref{eq:diff_def} by redefining the adjacency matrix as  $N A_{ij}\langle 0|i\rangle \langle j|0\rangle$, which reduces to $A_{ij}$ if $A$ is symmetric. Then, from Eq.~\eqref{eq:Lprime}, by defining 
$A'_{i'j'} \equiv N' f'_{i'j'} = -N'\langle 0|i'\rangle \langle j'|0\rangle L'_{i'j'}$ for $i'\neq j'$, we find $A'$ is related to $A$ by 
\begin{equation}
    A'_{i'j'} = {N'\over N}\sum_{i\in i', j\in j'} A_{ij}.
    \label{eq:Aprime}
\end{equation}
The connection strength between supernodes accounts for all connections between their member nodes while modulated by $N'/N$.  We use Eq.~\eqref{eq:Aprime} also for  $i'=j'$ to define $A'_{i'i'}$.  Consequently one can represent $L$ and $L'$ in Eq.~\eqref{eq:Lprime} also in terms of the adjacency matrices as  
\begin{equation}
   L_{ij} = \frac{q_{i}\delta_{ij} - A_{ij}}{N \langle 0|i\rangle \langle j|0\rangle} \to
   L'_{i'j'} = \frac{q'_{i'}\delta_{i'j'} - A'_{i'j'}}{N' \langle 0|i'\rangle \langle j'|0\rangle}
   \label{eq:Lform}
\end{equation}
with $q'_{i'} = \sum_{j'} A'_{j'i'} = \sum_{j'} A'_{i'j'}$.
Note that the mean connectivity is preserved,  $\bar{q}' \equiv \sum_{i'}q'_{i'}/N' = \bar{q}\equiv \sum_i q_i/N$, by Eq.~\eqref{eq:Aprime}. Note also that even if we begin with an unweighted adjacency matrix, the renormalized adjacency matrix generically becomes weighted.

The  \textit{equilibrium-preserving} LRG (EqLRG)  formulated in Eqs.~(\ref{eq:iprime}--\ref{eq:Aprime}) is the main contribution of this Letter. In the following, we apply this framework to hypergraphs to renormalize both elements and interactions, validating the EqLRG and gaining insights into how the structural characteristics of complex systems evolve across scales~\cite{Broido:2019aa,Serafino2021,MJLee2022,Bairey:2016aa,robiglio2024a}.

{\it Procedures and validation of EqLRG for hypergraphs.---}For a hypergraph in which a hyperedge $e\in \mathcal{E}$ with $E=|\mathcal{E}|$ may connect an arbitrary number of vertices $v\in \mathcal{V}$ with $V=|\mathcal{V}|$ and $B_{ve}\in \{0,1\}$ the  incidence matrix, we consider a simple diffusion process involving flows between a vertex and the hyperedge containing it~\cite{Zhou2007,Chitra2019,Carletti2020}. This can be described by Eq.~\eqref{eq:diff_def} in the bipartite-network representation, in which a node $i\in \mathcal{N} = \mathcal{V} \cup \mathcal{E}$ is a vertex or a hyperedge, and the adjacency matrix $A$ is given by $A_{ij}=B_{ve}$ if $\{i,j\}=\{v,e\}$ with $v\in \mathcal{V}$ and $e\in \mathcal{E}$  or $A_{ij}=0$ otherwise. The connectivity $q_i=\sum_{j}A_{ji}$ of node $i$ corresponds to the degree $k_v$ if $i=v\in \mathcal{V}$ or the cardinality $c_e$ if $i=e\in \mathcal{E}$. Note that projecting Eq.~\eqref{eq:diff_def} onto the vertex space, one can obtain a diffusion process generated by a Laplacian defined on the vertex space~\cite{Chitra2019}.  While multi-order~\cite{Cheng2024} and cross-order Laplacians~\cite{Nurisso2025}, accounting for diffusion over hyperedges of specific sizes separately, have been employed in LRG studies in simplicial complexes, we consider a simpler diffusion process to focus on the evolution of the effective structure sustaining diffusion  across scales. 

Vertices and hyperedges in local equilibrium are merged separately into supervertices and superhyperedges, respectively, preserving the bipartite structure. To reduce the number of  nodes by fraction $f$ at a timescale $t_c$, we iteratively connect the pairs of vertices ($i,j\in \mathcal{V}$) or hyperedges ($i,j\in \mathcal{E}$) in descending order of the propagator ratio 
\begin{equation}
\frac{\rho_{ij}(t_c)\rho_{ji}(t_c)}{\rho_{ii}(t_c)\rho_{jj}(t_c)}
\label{eq:merge_ij2}
\end{equation}
until the total number of connected components---each
representing a supervertex or a superhyperedge---is reduced to $N'=(1-f)N$. The time  $t_c$ is chosen to satisfy $S(t_c) = \ln ((1-  \kappa f)N)$, where the information entropy $S(t)\equiv -\mathrm{Tr} \left[\frac{\rho(t)}{Z(t)} \ln \frac{\rho(t)}{Z(t)}\right]$ with $Z(t)\equiv \text{Tr} \rho(t)$~\cite{Villegas:2023aa} represents the logarithm of the effective number of relevant eigenmodes, decreasing from $\ln N$ to $0$ as $t$ increases, and the factor $\kappa>1$ (set to $4$ in this study) ensures that the connected pairs are sufficiently equilibrated at $t_c$~\cite{SM}. After identifyng all supervertices and superhyperedges, the Laplacian $L'$ and the adjacency matrix $A'$ (or equivalently the incidence matrix $B'$) are constructed using Eqs.~\eqref{eq:Lprime} and \eqref{eq:Aprime}. Time is rescaled by Eq.~\eqref{eq:tprime}. Repeating this process generates hypergraphs at coarser scales. 

\begin{figure}
    \centering
    \includegraphics[width=0.48\textwidth]{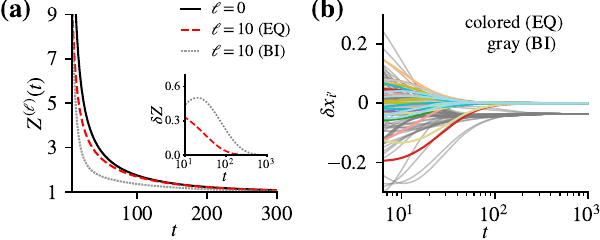}
    \caption{Validation of the renormalized diffusion for a BA hypertree with $V=100$ vertices and $E=100$ hyperedges under EqLRG (EQ), compared with the binary (BI) scheme. (a) Traces of the original diffusion propagator $Z^{(0)}(t)$ (solid) and the renormalized one $Z^{(\ell)}(t')$ (dashed) at the $\ell=10$ RG step with $f=0.1$ and $t'$ related to $t$ by Eq.~\eqref{eq:tprime}. Inset: Plot of  the relative error $\delta Z(t) \equiv \frac{Z^{(0)}(t)-Z^{(10)}(t')}{Z^{(0)}(t)}$ versus time $t$.
    (b) Relative error of the solutions $x'_{i'}(t')$'s of Eq.~\eqref{eq:diff_rg} at the $\ell=10$ RG step with respect to the coarse-grained original solutions $x_{i'}(t)=\sum_{i\in i'}\langle i^{'}|i\rangle x_i(t)$ for EQ and $x_{i'}(t)=\sum_{i\in i'}x_i(t)/\sum_{i\in i^{'}}1$ for BI ~\cite{SM}.
    The error is quantified as $\delta x_{i'}(t) \equiv \frac{x^{'}_{i'}(t') - x_{i'}(t)}{x_{i'}(t)}$, evaluated  for all supernodes $i'$ in a single diffusion realization with a random initial condition. 
    }
    \label{fig:traj}
\end{figure}

\begin{figure*}
    \centering
    \includegraphics[width=\textwidth]{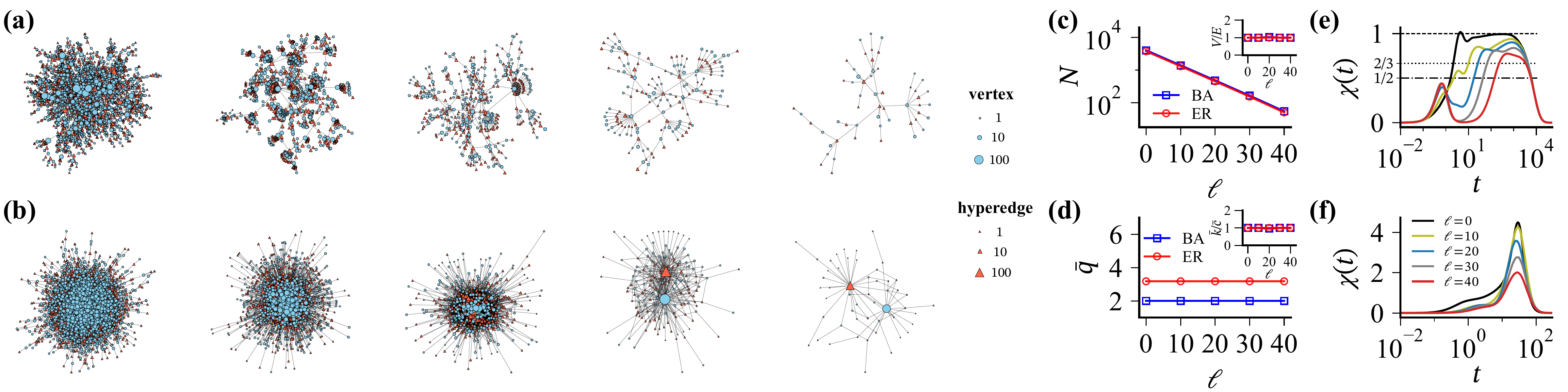}
    \caption{
    Transformation of hypergraphs by EqLRG.
    (a) Transformation of a BA hypertree under recursive EqLRG with $f=0.1$ from the original one ($\ell=0$) to the transformed ones at RG steps $\ell=10,20,30$ and $40$ (from left to right). They are  represented as bipartite networks, where circles and triangles represent vertices and hyperedges, with sizes increasing with degree and cardinality, respectively. 
    (b) Same as (a) for the giant component of an ER hypergraph ($\bar{k}=\bar{c} =3$) under the EqLRG with $f=0.1$. 
    (c) Number of supernodes ($N$) for BA and ER  versus RG steps $\ell$. Inset: The ratio of the numbers of supervertices ($V$)  and superhyperedges ($E$). 
    (d) Mean connectivity $\bar{q}$ is preserved under the EqLRG. Inset: The ratio of the mean vertex degree to the mean hyperedge cardinality $\bar{k}/\bar{c}$.
    Heat capacity $\chi(t)$ vs.\ time $t$ at different RG steps $\ell$ for (e) BA and (f) ER.
    Points and errorbars denote averages and standard deviations over $10^3$ realizations. $V=E=2\times 10^3$ in both cases. 
    }
    \label{fig:trans}
\end{figure*}

We consider various hypergraphs extended from their network counterparts~\cite{Barabasi1999_BA,Villegas:2023aa, Goh2001_Static}; (i) Barab\'{a}si-Albert-type (BA)  hypertrees with both degree and cardinality following power-law distributions $P(k)\sim k^{-3}$ and $P(c)\sim c^{-3}$; (ii) Random (RN) hypertrees exhibiting Poissonian degree and cardinality distributions; (iii) Erd\H{o}s-R\'{e}nyi (ER) hypergraphs characterized by Poisson degree and cardinality distributions with mean values $\bar{k}=\bar{c}=3$~\cite{SM}. The two hypertrees are {\it informationally scale-invariant} in that  the entropy decays as $S(t) \simeq \ln N - \frac{d_{\rm s}}{2}\ln t$ and the ``heat capacity" $\chi \equiv -\frac{dS}{d\ln t}$ takes a constant value $\chi=\frac{d_{\rm s}}{2}$ over a wide time range with the spectral dimension $d_{\rm s}=2$ and $d_{\rm s}=\frac{4}{3}$ for BA and RN hypertrees, respectively. In contrast, ER hypergraphs do not exhibit such plateau in $\chi(t)$, indicating the absence of a well-defined spectral dimension~\cite{Villegas:2023aa,poggialini2024}.

The traces of diffusion propagators,  $Z(t)= \sum_{\lambda}e^{-\lambda t}$ and $Z'(t')=\sum_{\lambda'} e^{-\lambda' t'}$,  corresponding to the Laplace transforms of the spectral density functions of the original and renormalized Laplacian, respectively, closely match for large $t$ and $t'$  related by Eq.~\eqref{eq:tprime}  [Fig.~\ref{fig:traj}(a)]. This confirms the preservation of small Laplacian eigenvalues throughout the EqLRG. Moreover, the solutions $x'$ of Eq.~\eqref{eq:diff_rg} converge to those of Eq.~\eqref{eq:diff_def} in the long-time limit [Fig.~\ref{fig:traj}(b)]. The formulation of 
$L'$ in Eq.~\eqref{eq:Lprime} plays a crucial role in ensuring this convergence. For comparison, we consider an alternative renormalization scheme: $L^{\prime \ {\rm (BI)}}_{i'j'} \equiv q_{i'}^{\rm (BI)} \delta_{i'j'} - A^{\prime {\rm (BI)}}_{i'j'}$ where binary connection strength is enforced as $A^{\prime {\rm (BI)}}_{i'j'} = \theta(\sum_{i\in i', j\in j'} A_{ij})$ with $\theta(x>0)=1$ and $\theta(x\leq 0)=0$~\cite{SM}. In Fig.~\ref{fig:traj}, $Z'(t')$ constructed by $L^{\prime \ {\rm (BI)}}$ shows larger deviations from $Z(t)$ for large $t'$ and the solutions to Eq.~\eqref{eq:diff_rg} with $L^{\prime \ {\rm (BI)}}$ deviate significantly from the original solutions, demonstrating it is crucial to construct the correct renormalized $L'$ by Eq.~\eqref{eq:Lprime} to preserve the equilibrium state.  

The transformation of a BA hypertree and an ER hypergraph under recursive EqLRG is shown in Fig.~\ref{fig:trans}(a,b). The BA hypertree appears to undergo moderate changes, retaining tree structure, whereas the ER hypergraph evolves significantly, developing hub vertices and hyperedges and eventually forming a single ultra-hub vertex and hyperedge in  the final stages when only a few supernodes remain. The number of nodes $N=V+E$ decays exponentially with RG step $\ell$ as $\sim (1-f)^\ell$ [Fig.~\ref{fig:trans}(c)] while the mean connectivity $\bar{q}$ is preserved [Fig.~\ref{fig:trans}(d)].

{\it Evolution of spectral dimensions and connectivity structure.---}For BA hypertrees, the initial heat capacity plateau at $\chi=1$ gradually shifts toward $\chi= \frac{2}{3}$ as the RG progresses while developing a peak, approaching $\chi=\frac{1}{2}$, at early times [Fig.~\ref{fig:trans}(e)]. RN hypertrees maintain the $\chi=\frac{2}{3}$-plateau while also developing a peak at early times~\cite{SM}. These results suggest that the effective dimensions of both hypertrees converge to an effective spectral dimension $d_s=\frac{4}{3}$ at large spatiotemporal scales. In contrast, ER hypergraphs  exhibit only a sharp peak in $\chi(t)$ at all RG steps, indicating that their spectral dimension remains undefined or effectively infinite [Fig.~\ref{fig:trans}(f)].

\begin{figure}
    \centering
    \includegraphics[width=0.48\textwidth]{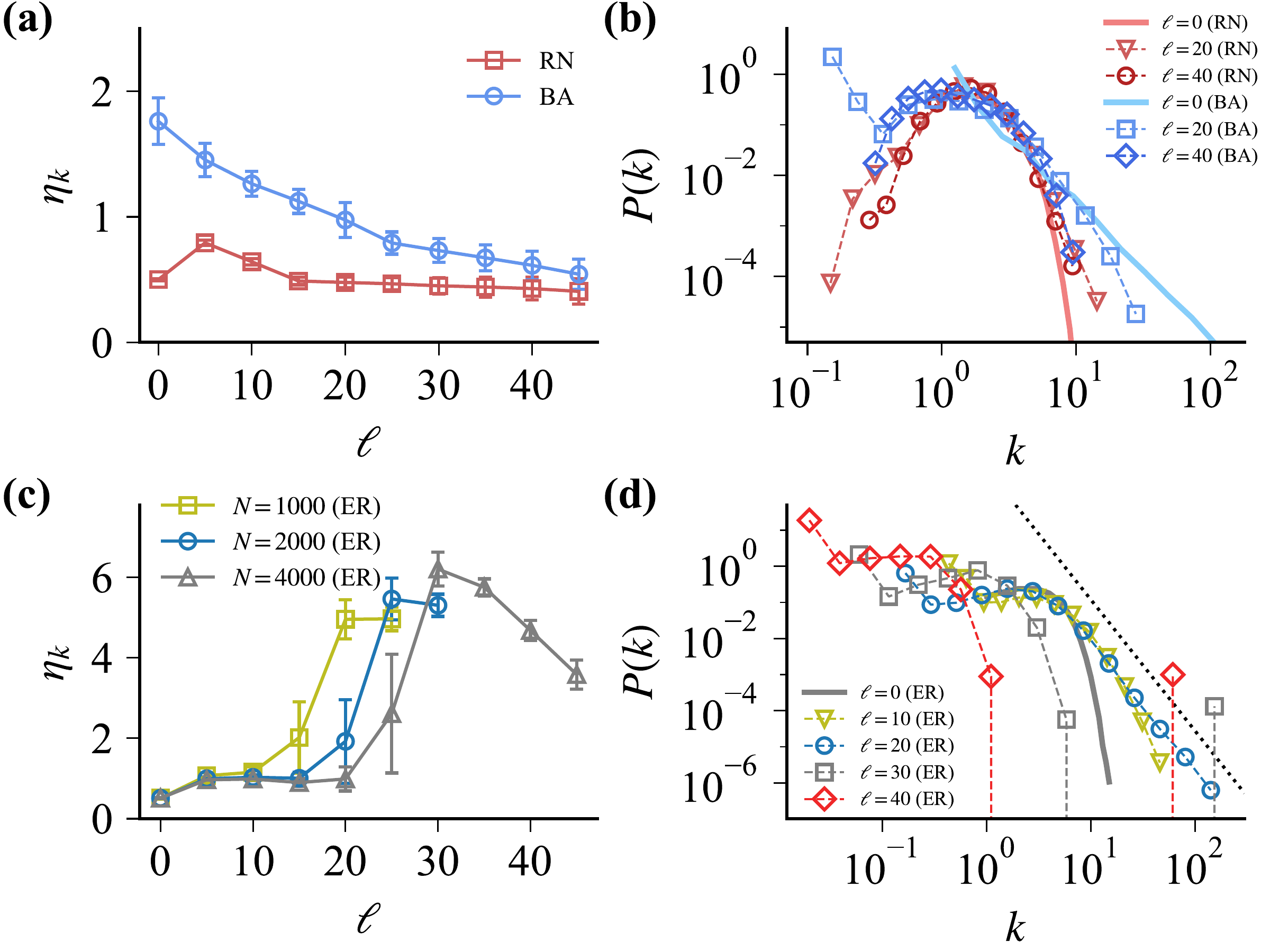}
    \caption{
    Evolution of connectivity structure under EqLRG.
    (a,b) Results from $10^3$ realizations of RN and BA hypertrees and (c,d) for ER hypergraphs.  Same hypergraph parameters are used and $f=0.1$ as in Fig.~\ref{fig:trans}.
    (a,c) Degree heterogeneity index $\eta_k$ versus RG steps $\ell$, with errorbars denoting standard deviations.     (b,d) Degree distribution $P(k)$ at various RG steps $\ell$. In (b), RN and BA converge to  similar Poissonian forms at large $\ell$, while in (d), ER hypergraphs develop power-law behavior  at $\ell=10$ and $20$; a dotted line with slope $-4$ is shown as a visual guide. 
    }
    \label{fig:dist}
\end{figure}

To explore the evolution of connectivity structure, we introduce the degree heterogeneity index $\eta_k\equiv \frac{\sqrt{\bar{k^2} - \bar{k}^2}}{\bar{k}}$. We here omit $\eta_c$ as it exhibits the same behavior as $\eta_k$ in the considered hypergraphs~\cite{SM}. As the RG progresses, the values of $\eta_k$ for both RN and BA hypertrees remain near or converge toward a value close to $\frac{1}{2}$ [Fig.~\ref{fig:dist}(a)], matching that of the original RN hypertrees~\cite{SM}. Their degree distributions also remain close to or evolve toward the Poissonian form of the original RN hypertrees [Fig.~\ref{fig:dist}(b)]. These results suggest that the structure of RN hypertrees---specifically their spectral dimension and connectivity distributions---is dynamically scale-invariant and attractive in that they emerge to support late-time diffusion dynamics in systems with low spectral dimensions.
On the other hand, BA hypertree is informationally and structurally scale-invariant~\cite{Villegas:2023aa,poggialini2024} but it is not dynamically scale-invariant, suggesting decoupling of structural and dynamical scale-invariance.

ER hypergraphs evolve in different manners. The degree heterogeneity first increases and remain constant before rising sharply in the final stages due to the formation of ultra-hubs [Fig.~\ref{fig:dist}(c)]. The degree distribution broadens, evolving first from Poisson to power-law and then humps are developed in the final stages [Fig.~\ref{fig:dist}(d)]. The intermediate RG regime exhibiting a constant $\eta_k$ and power-law degree distributions extends with system size, suggesting that the effective structure for diffusion at larger scales in ER hypergraphs is characterized by  high diversity of degrees and cardinality, despite the intial narrow distributions.

EqLRG reveals distinct effective structures for late-time diffusion between 
informationally scale-invariant (finite $d_s$) versus scale-variant (undefined $d_s$) hypergraphs: RN-hypertree-like structures emerge in the former while heterogeneous connectivity structures emerge in the latter. The emergence of such  a narrow or broad spectrum of degrees and cardinalities at long timescales may be related to the nature of the supernode formation and growth driven by the spread of local equilibrium over time. As time scale increases, more and more node pairs reach local equilibrium to form and grow supernodes. It turns out that the cluster-size of a supernode---defined as the number of nodes within each supernode---follows a narrow or power-law-like distribution  mirroring the connectivity  distributions of the supernodes~\cite{SM}.  This parallels the mass aggregation process with input~\cite{Krapivskybook2010}, for which the role of the effective spectral dimension in shaping the cluster-size distribution may be investigated. The emergence of heterogeneity at coarser scales in ER hypergraphs suggests that the ubiquity of heterogeneous and higher-order-interaction networks at currently available empirical resolutions may, in part, result from  the coarse-graining of originally more homogeneous and lower-order-interaction structures at finer scales. 

{\it Discussion.---}Leveraging the notion that slow eigenmodes of Laplacian matrix remain representable by a proper, quasi-complete basis after merging nodes in local equilibrium, we have formulated  in Eqs.~(\ref{eq:iprime}--\ref{eq:Aprime}) a rigorous LRG scheme preserving the diffusion dynamics at long timescales, that we termed the equilibrium-preserving LRG (EqLRG), and a principled coarse-graining procedure by aggregation of equilibrium-state flows. By applying this EqLRG to various hypergraphs, we observed distinct structural evolutions of dynamically scale-invariant systems depending on the spectral dimensionality, and discussed its origin and implications in modeling complex systems at finite empirical resolutions. Juxtaposing our results with those of original LRG~\cite{Villegas:2023aa,poggialini2024} indicates decoupling of structural and dynamical scale-invariance.

There are several future directions toward applying and developing further our framework to more complex and empirical systems. Studying the RG evolution of structural features beyond heterogeneity, such as clustering, community structure~\cite{Villegas2025}, and structural correlations~\cite{LeeMJ2024}, can offer insights into multi-scale properties. Further investigation of the growth and merging of supernodes may clarify the origin of the structural evolution with changing resolution. The EqLRG framework can also be extended to incorporate generalized diffusion processes with rates dependent on degree, cardinality, and higher-order structure~\cite{KimJ2023,KimJH2024}. Finally, coarse-graining real-world systems using EqLRG~\cite{SM} can practically reduce complexity while preserving late-time dynamics, offering opportunities toward systematic analysis of effective dynamics across scales.

\textit{Acknowledgments.}--- This work was supported in part by the National Research Foundation of Korea (NRF) grant funded by the Korea government (MSIT) [No.~RS-2022-NR071795 and No.~RS-2023-00279802 (S.Y.); No.~RS-2025-00558837 (K.-I.G.)], by KIAS Individual Grants [Nos.~CG074102 (S.Y.) and CG079902 (D.-S.L.)] at Korea Institute for Advanced Study, and by the KENTECH Research Grant [No.~KRG-2021-01-007 (S.Y.)]. We thank the Center for Advanced Computation in KIAS for providing the computing resources.

\bibliography{references}

\end{document}


\title{Equilibrium-preserving Laplacian renormalization group:\\ Supplemental Material}
\author{Sudo  \surname{Yi}}
\affiliation{School of Computational Sciences, Korea Institute for Advanced Study, Seoul 02455, Korea}
\affiliation{CCSS, KI for Grid Modernization, Korea Institute of Energy Technology, 21 Kentech, Naju, Jeonnam 58330, Korea}
\author{Seong-Gyu  \surname{Yang}}
\affiliation{School of Computational Sciences, Korea Institute for Advanced Study, Seoul 02455, Korea}
\affiliation{Integrated Science Lab, Department of Physics, Ume\r{a} University, SE-901 87, Sweden}
\author{K.-I. \surname{Goh}}
\email{kgoh@korea.ac.kr}
\affiliation{Department of Physics, Korea University, Seoul 02841, Korea}
\author{D.-S. \surname{Lee}}
\email{deoksunlee@kias.re.kr}
\affiliation{School of Computational Sciences, Korea Institute for Advanced Study, Seoul 02455, Korea}
\affiliation{Center for AI and Natural Sciences, Korea Institute for Advanced Study, Seoul 02455, Korea}
\date{\today}

\maketitle
\tableofcontents

\section{Derivation of EqLRG}

In this section, we provide the background and a detailed derivation of the equilibrium-preserving Laplacian Renormalization Group (EqLRG) method.

\subsection{Spectral-space and real-space LRG} 
\label{sec:spectral}

Here we first review the theoretical background of the Laplacian Renormalization Group (LRG) introduced in Ref.~\cite{Villegas:2023aa}. Let us consider a diffusion process for a network with $\mathcal{N}$ the set of nodes and $A$ the adjacency matrix. Adopting the bra-ket notation, we let $|x(t)\rangle$ denote the dynamical state of the considered system at time $t$ and $L$ the Laplacian operator generating the diffusion dynamics. The diffusion equation is given  in the operator form as
\begin{equation}
    \frac{\partial}{\partial t} |x(t)\rangle = - L |x(t)\rangle.
    \label{eq:diff_op}
\end{equation}
For an initial condition $|x(0)\rangle$,  the state at time $t$ is given by  $|x(t)\rangle = \rho(t)|x(0)\rangle$ with the unnormalized propagator given by
\begin{equation}
    \rho(t) \equiv e^{-Lt}= \sum_\lambda e^{-\lambda t} |\lambda\rangle \langle \lambda|
    \label{eq:rho_def}
\end{equation} 
with $|\lambda\rangle (\langle \lambda|)$ denoting the right(left) eigenvector associated with the eigenvalue $\lambda$ of the Laplacian $L$. Let $\lambda_1, \lambda_2, \ldots, \lambda_N$ denote the eigenvalues of the Laplacian sorted in the ascending order such that $\lambda_1<\lambda_2\leq \lambda_3\leq \cdots \leq \lambda_N$, where $N=|\mathcal{N}|$. The Laplacian is positive semi-definite, i.e., all $\lambda$'s are non-negative, and the smallest eigenvalue is zero, $\lambda_1=0$. The trace of the propagator 
\begin{equation}
    Z(t) \equiv \textrm{Tr}\rho = \sum_\lambda e^{-\lambda t}
    \label{eq:Ztdef}
\end{equation}
corresponds (up to a constant factor $N$) to the Laplace transform of the spectral density function 
\begin{equation}
p(\lambda) \equiv N^{-1} \sum_n \delta(\lambda_n - \lambda).
\label{eq:pdef}
\end{equation}

In the late-time regime $t\gg t_c$ with a given timescale $t_c$, fast modes, i.e., the eigenmodes with as large eigenvalues as $\lambda t_c> 1$, make negligible contributions to the propagator  since $e^{-\lambda t}\ll 1$ for $t\gg t_c$. Therefore, we have an approximation for Eq.~\eqref{eq:rho_def} as 
\begin{equation}
    \rho(t \gg t_c) \approx \sum_{\lambda< t_c^{-1}} e^{-\lambda t} |\lambda\rangle \langle \lambda|,
    \label{eq:rho_approx}
\end{equation} 
and 
\begin{equation}
    Z(t\gg t_c) \approx \sum_{\lambda<t_c^{-1}} e^{-\lambda t},
\end{equation}
which indicates a reduced number of relevant eigenmodes and thus we call the spectral-space LRG~\cite{Villegas:2023aa}.

To consider the real-space LRG, let us first recall that in the node basis $\{|i\rangle | i\in \mathcal{N} \}$  and its dual $\{\langle i|\}$, the dynamical state  and the Laplacian  operator are represented as $|x(t)\rangle = \sum_{i\in \mathcal{N}} |i\rangle \langle i|x(t)\rangle = \sum_{i\in \mathcal{N}} |i\rangle  x_i(t)$ and $L = \sum_{i,j\in \mathcal{N}} |i\rangle \langle i|L|j\rangle \langle j|= \sum_{i,j\in \mathcal{N}} |i\rangle L_{ij} \langle j|$, due to the completeness of the node basis $\langle i|j\rangle = \delta_{ij}$, and we obtain Eq. (1) of the main text as the {\it real} space representation of Eq.~\eqref{eq:diff_op}. Also the propagator is represented in this basis as 
\begin{equation}
\rho(t) = \sum_{i\in \mathcal{N}, j\in \mathcal{N}} |i\rangle \rho_{ij}(t) \langle j| \quad {\rm  with} \quad  \rho_{ij}(t) =\sum_\lambda e^{-\lambda t} \langle i|\lambda\rangle \langle \lambda |j\rangle.
\label{eq:rhoij_def}
\end{equation}
We use $i$ and $j$ to denote nodes in the set of nodes $\mathcal{N}$ unless stated otherwise in the present study.  Then, similarly to the spectral-space LRG in Eq.~\eqref{eq:rho_approx}, where the late-time diffusion dynamics is approximated by a reduced number of slow eigenmodes,  one may consider a complementary approach in the real space, in which  the diffusion dynamics is represented  with a reduced set of effective block-nodes or {\it supernodes}~\cite{Villegas:2023aa}. In this real-space LRG, two nodes $i$ and $j$ are merged into the same supernode if they satisfy the condition~\cite{Villegas:2023aa}
\begin{equation}
\frac{\rho_{ij}(t_c)}{\min (\rho_{ii}(t_c), \rho_{jj}(t_c))}\geq 1,
\label{eq:merge_ij}
\end{equation}
based on the idea that they are {\it proximate} on the timescale $t_c$ if diffusion from $j$ to $i$ quantified by the propagator $\rho_{ij}(t)$ is comparable to self-diffusion, $\rho_{ii}(t)$ or $\rho_{jj}(t)$.  The resulting network of supernodes serves as an effective  representation of late-time diffusion, providing a coarse-grained network of the  original one with resolution controlled by $t_c$. The diffusion time is used as a resolution parameter to probe the global organization of a given system, enabling validation of the effective underlying structure. In this way, the LRG approach may advance the renormalization-group methods for investigating the structure and dynamics of networks. 

However, the real-space LRG framework still lacks a fully systematic foundation. For example, the rationale behind using Eq.~\eqref{eq:merge_ij} to assess node proximity and whether the reduced set of block-nodes can reliably capture the original dynamics remain unanswered. The latter is essential for validating the proposed LRG framework beyond structural perspective. Establishing this theoretical foundation is essential for extending the method to a broader class of networks and hypergraphs without ambiguity or risk of misinterpretation. In the following, we begin by formulating the condition for a reduced basis, i.e.,  a set of supernodes, to accurately approximate  late-time diffusion dynamics. We then  show that this condition naturally leads to a proper basis transformation from the original to supernode representation, that allows the renormalized diffusion equation to preserve the original long-time behavior. 

\subsection{Implications of Eq. (2) and derivation of Eq. (3)} 
\label{sec:Eqs23}
Suppose that the original node set $\mathcal{N} = \{i\}$ is partitioned into a set of supernodes $\mathcal{N}' = \{i'\}$, where each original node $i$ in $\mathcal{N}$ belongs to exactly one supernode $i'\in \mathcal{N}'$.  If the diffusion propagator $\rho(t)$ from Eq.~\eqref{eq:rho_def} is well represented in the supernode basis for $t\gg t_c$ as
\begin{equation}
    \rho(t\gg t_c) =e^{-Lt} \approx \sum_{i',j'\in \mathcal{N}'} |i'\rangle \langle i'|e^{-Lt}|j'\rangle \langle j'| = \sum_{i',j'\in \mathcal{N}'} |i'\rangle  \rho_{i'j'}(t)\langle j'|
    \label{eq:fullrep}
\end{equation}
with $\rho_{i'j'}(t)\equiv \langle i'|e^{-Lt}|j'\rangle= \sum_{i,j\in \mathcal{N}}\langle i'|i\rangle \langle i|e^{-Lt}|j\rangle \langle j|j'\rangle$, then the late-time diffusion dynamics can be reliably captured by the supernodes. The basis transformation coefficients $\langle i'|i\rangle$ and $\langle j|j'\rangle$ are derived in Eq. (4) in the main text and in details in the next subsection. Since $\rho(t)$ is governed by the eigenvectors corresponding to eigenvalues $\lambda<t_c^{-1}$ of the Laplacian, Eq.~(2) of the main text - which states that the slow eigenvectors are approximately representable by the supernode basis - leads to Eq.~\eqref{eq:fullrep}. 
Note that Eq.~\eqref{eq:fullrep} can be considered as the real-space LRG compared to Eq.~\eqref{eq:rho_approx}.

\begin{figure}[h]
\includegraphics[width=0.6\columnwidth]{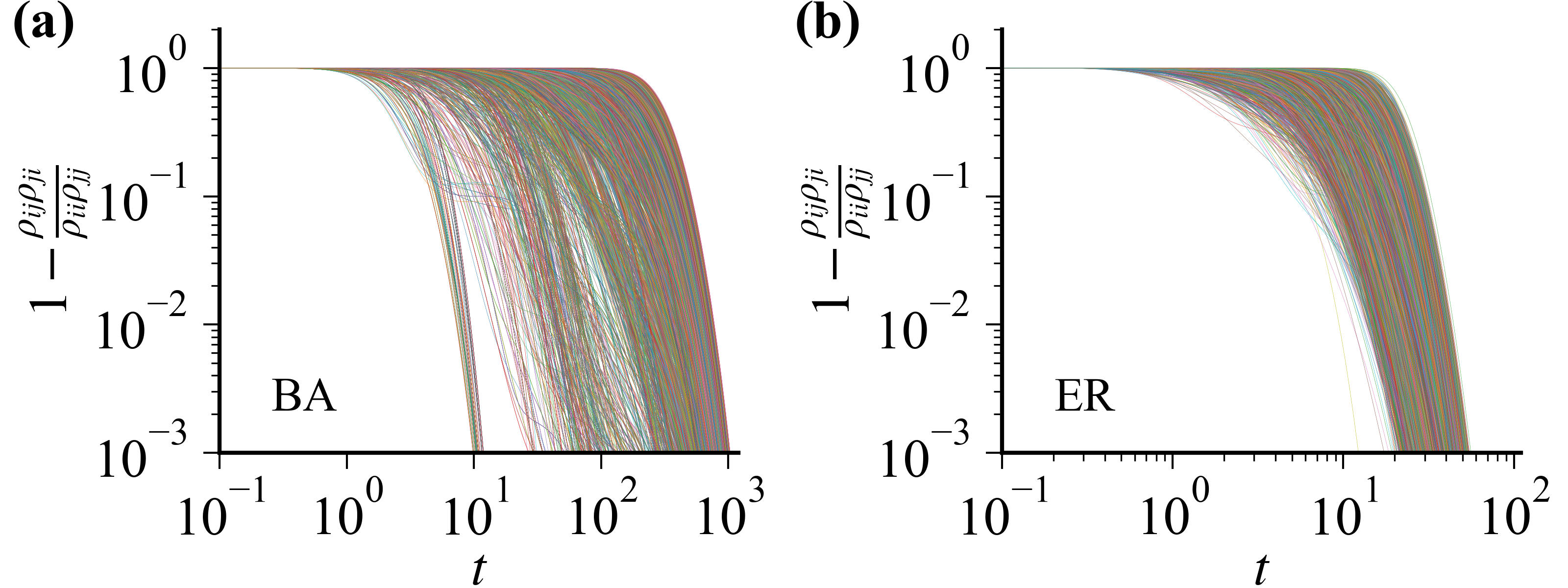}
\caption{Convergence of the propagator ratio $\frac{\rho_{ij}(t) \rho_{ji}(t)}{\rho_{ii}(t)\rho_{jj}(t)}$ to $1$ for all pairs of nodes $i$ and $j$ in (a) a BA hypertree  and (b) the giant component of a ER hypergraph with $\bar{k} =3$ and  $\bar{c} =3$. Different pairs exhibit varying convergence rates.}
\label{fig:ratio}
\end{figure}

The quasi-completeness of the supernode basis, as expressed in Eq.~(2) of the main text or equivalently in Eq.~\eqref{eq:fullrep},  provides insight into the physical meaning of supernodes - namely, that they represent regions of {\it local equilibrium}. Specifically, original nodes $i$ and $j$ belonging to the same supernode $i'$ exhibit similar propagator profiles for $t\gg t_c$, such that $\rho_{ik}(t)\propto \rho_{jk}(t)$ with a proportionality constant independent of $k$ or $t$. This follows from
\begin{equation} 
\rho_{ik}(t\gg t_c) \approx \sum_{\lambda<t_c^{-1}} \langle i|\lambda \rangle \langle \lambda|k\rangle e^{-\lambda t}\approx \sum_{\lambda<t_c^{-1}} \langle i|i'\rangle \langle i'|\lambda \rangle \langle \lambda|k\rangle e^{-\lambda t} \approx \frac{\langle i|i'\rangle}{\langle j|i'\rangle} \rho_{jk}(t\gg t_c),
\label{eq:rhoratio}
\end{equation}
using Eq.~(2). Hence, the ratio $\frac{\rho_{ik}(t)}{\rho_{jk}(t)}$ becomes constant for $t\gg t_c$, indicating that $i$ and $j$ are in local equilibrium. This mirrors the behavior in global equilibrium, where for all node pairs $i$ and $j$,   the ratio becomes $\frac{\langle i|0\rangle}{\langle j|0\rangle}$ with $|0\rangle$ and $\langle 0|$ the right and left eigenvectors associated with the zero eigenvalue $\lambda_1=0$ of the Laplacian. Furthermore, Eq.~\eqref{eq:rhoratio} implies 
\begin{equation}
\frac{\rho_{ij}(t) \rho_{ji}(t)}{\rho_{ii}(t)\rho_{jj}(t)}  = \frac{\rho_{ij}(t)}{\rho_{jj}(t)}\frac{\rho_{ji}(t)}{\rho_{ii}(t)} \approx \frac{\langle i|i'\rangle}{\langle j|j'\rangle} \frac{\langle j|i'\rangle}{\langle i|i'\rangle}=1, 
\label{eq:rho4}
\end{equation}
which converges to $1$ at different rates for different node pairs as shown in Fig.~\ref{fig:ratio}. This convergence can thus be used as a practical criterion to identify local equilibrium pairs. Equation~\eqref{eq:merge_ij}, employed in Ref.~\cite{Villegas:2023aa}, can be seen as  a relaxed condition of Eq.~\eqref{eq:rho4} for determining whether two nodes should be assigned to the same supernode at a given time scale $t_c$.

By using the renormalized propagator $\rho_{i'j'}(t)$ in Eq.~\eqref{eq:fullrep} in the late-time regime ($t\gg t_c$), one can represent  the dynamical state $|x(t)\rangle = \rho(t)|x(0)\rangle $ approximately in the supernode basis as $|x(t)\rangle \approx \sum_{i'\in \mathcal{N}'} |i'\rangle x_{i'}(t)$ with the coarse-grained variables $ x_{i'}(t)\equiv \langle i'|x(t)\rangle = \langle i'|\rho(t) |x(0)\rangle$.  Inserting it into Eq.~\eqref{eq:diff_op}, we obtain
\begin{equation}
    \frac{\partial}{\partial t} \sum_{i'\in \mathcal{N}'} |i'\rangle \langle i'|x(t)\rangle \approx -L\sum_{j'\in \mathcal{N}'} |j'\rangle \langle j'|x(t)\rangle,
    \label{eq:diff_med}
\end{equation}
and applying $\langle i'|$, we find the coarse-grained variables satisfy approximately  
\begin{equation}
    \frac{\partial}{\partial t} x_{i'}(t) \approx -\sum_{j'\in \mathcal{N}'} \langle i'|L|j'\rangle x_{j'}(t)
    \label{eq:xcg}
\end{equation} 
Here we introduce the renormalization Laplacian 
\begin{equation}
    L'_{i'j'}\equiv \langle i'|L|j'\rangle = \sum_{i,j\in \mathcal{N}}\langle i'|i\rangle \langle i|L|j\rangle \langle j|j'\rangle.
\end{equation} 
Note that the coarse-grained variables $x_{i'}(t)$ satisfy only approximately Eq.~\eqref{eq:xcg} since $| i'\rangle\langle i'|$ is identity only when acted upon the slow eigenmodes of the Laplacian. Let us introduce another dynamical state $|x'\rangle$ generated by the renormalized Laplacian $L'$ spanned in the supernode basis as 
\begin{equation}
    \frac{\partial}{\partial t'} |x'(t')\rangle = -L'|x'(t')\rangle
        \label{eq:diff_rg_op}
\end{equation}
with a rescaled time $t'$. The solution is then given by 
\begin{equation}
    |x'(t')\rangle = \rho'(t')|x'(0)\rangle  \quad {\rm with} \quad \rho'(t') \equiv e^{-L't'} = \sum_{\lambda'} e^{-\lambda' t'} |\lambda'\rangle \langle \lambda'|,
\end{equation}
where $\lambda'$ and $|\lambda'\rangle (\langle \lambda'|)$ are the eigenvalues and the right (left) eigenvectors of the renormalized Laplacian spanned in the supernode basis. Eq. (3) of the main text is the representation of Eq.~\eqref{eq:diff_rg_op} in the supernode basis. Assuming the quasi-completeness of the supernode basis in Eq. (2), one can expect that the spectra of the renormalized Laplacian remain close to those of the original Laplacian and that the coarse-grained variable $x_{i'}(t) = \langle i'|x(t)\rangle = \langle i'|\rho(t)|x(0)\rangle$ from the original solution to Eq. (1) or Eq.~\eqref{eq:diff_op} is well approximated by the solution to Eq. (3) or Eq.~\eqref{eq:diff_rg_op} as 
\begin{equation}
   x_{i'}(t) = \langle i'|\rho(t)|x(0)\rangle =\sum_{i\in i'} \langle i'|i\rangle x_i(t)\approx  x'_{i'}(t') = \langle i'|\rho'(t')|x'(0)\rangle.
    \label{eq:xprimex}
\end{equation}
Note that $\rho(t)$ in Eq.~\eqref{eq:fullrep} is not identical to $\rho'(t')$ although both are represented in the supernode basis. Moreover, to ensure the approximation in Eq.~\eqref{eq:xprimex}, we relate the rescaled time $t'$ to the original time $t$ by Eq. (6) of the main text. 

The similarity of the spectra of the original and the renormalized Laplacian and the validity of Eq.~\eqref{eq:xprimex} are confirmed in Fig. 1 in the main text and will further be analyzed in Sec.~\ref{sec:accuracy}.

\subsection{Basis transformation and normalization in  Eq. (4)} 
\label{sec:Eq4}

In the main text, we derived the basis transformation $\langle i'|i\rangle$ and $\langle j|j'\rangle$ in Eq. (4) by equating the constant propagator ratios $\frac{\rho_{ik}(t)}{\rho_{ik}(t)}$ in local and global equilibrium. This reveals the physical meaning of supernodes. Also the preservation of the eigenvectors corresponding to the zero eigenvalue $|0\rangle$ and $\langle 0|$ under the EqLRG follows directly from the formulation in Eq. (4). 

Conversely, we can derive Eq. (4) by assuming, instead, that the eigenvectors for $\lambda=0$ are exactly represented in  the supernode basis---not merely approximately. To see this, suppose that Eq. (2) in the main text holds exactly for $\lambda = 0$, i.e., 
\begin{equation}
 |0\rangle = \sum_i |i\rangle \langle i|0\rangle = \sum_{i'} |i'\rangle \langle i'|0\rangle \quad {\rm and} \quad \langle 0|=\sum_i \langle 0|i\rangle \langle i| = \sum_{i'} \langle 0|i'\rangle \langle i'|
 \label{eq:00}
\end{equation}
Applying $\langle i|$ and $|i\rangle$ to Eq.~\eqref{eq:00}, we obtain,  for each node $i$ belonging to supernode $i'$, 
\begin{equation}
    \langle i|0\rangle = \langle i|i'\rangle \langle i'|0\rangle \ {\rm and} \ \langle 0|i\rangle = \langle 0|i'\rangle \langle i'|i\rangle
\end{equation}
which coincides with Eq. (4) in the main text. This implies that the components $\langle i|i'\rangle$ of nodes within a supernode $i'$ are proportional to their components in the  eigenvector, i.e., $\langle i|i'\rangle \propto \langle i|0\rangle$,  their local equilibrium. The invariance of the left eigenvector,  $\langle 0'| =\langle 0|$,  leads the conservation law, stating that $\langle 0|x(t)\rangle = \sum_i \langle 0|i\rangle x_i(t)$ is constant over time,  to be valid also in the supernode basis, where $\langle 0|x'(t')\rangle = \sum_{i'} \langle 0|i'\rangle x'_{i'}(t')$ is constant. 

The basis vectors $|i'\rangle$ and $\langle i'|$ in Eq. (4) are required to satisfy the normalization condition  $\langle i'|i'\rangle = 1$. Consequently the two coefficients $\langle i'|0\rangle$ and $\langle 0|i'\rangle$ in Eq. (4) should satisfy
\begin{equation}
    \langle i'|0\rangle \langle 0|i'\rangle = \phi_{i'} \equiv \sum_{i\in i'} \langle i|0\rangle \langle 0|i\rangle
    \label{eq:norm}
\end{equation} 
where $\phi_{i'}$ can be interpreted as the relative weight of supernode $i'$ in the eigenvector for $\lambda'=0$ and satisfies $\sum_{i'}\phi_{i'} = 1$. Also one can define $\phi_i\equiv \langle i|0\rangle \langle 0|i\rangle$, which  represents the weight of node $i$ in the $\lambda=0$ eigenvector of the original Laplacian.  This allows Eq.~\eqref{eq:norm} to be rewritten as $\phi_{i'} = \sum_{i\in i'} \phi_i$.  For initial networks with Laplacian $L_{ij} = q_i \delta_{ij} - A_{ij}$ with $A_{ij}=0,1$ and $q_i = \sum_j A_{ij}$, we have  $\phi_i = N^{-1}$  As long as Eq.~\eqref{eq:norm} is satisfied, the magnitudes of the basis vectors can be freely chosen.  

In this work, we adopt a symmetric normalization for both the original and the supernode basis vectors as $\langle 0|i\rangle = \langle i|0\rangle = \sqrt{\phi_{i}}$ and $\langle 0|i'\rangle = \langle i'|0\rangle = \sqrt{\phi_{i'}}$.  Under this scheme, the basis vectors for a supernode $i'$ in Eq. (4) in the main text are given by
\begin{equation}
|i'\rangle = \sum_{i\in i'} |i\rangle \sqrt{\frac{\phi_i}{\phi_{i'}}} \ {\rm and} \ 
\langle i'| = \sum_{i\in i'} \sqrt{\frac{\phi_i}{\phi_{i'}}} \langle i|.
\end{equation}
Accordingly the coarse-grained variable $x_{i'}(t)$ at supernode $i'$ is given by 
\begin{equation}
    x_{i'}(t) = \langle i'|x(t)\rangle = \sum_{i\in i'} \langle i'|i\rangle \langle i|x(t)\rangle = \sum_{i\in i'} \sqrt{\frac{\phi_i}{\phi_{i'}}} x_i(t),
\end{equation}
and the renormalized Laplacian is given by
\begin{equation}
    L'_{i'j'} = \sum_{i\in i', j\in j'} \sqrt{\frac{\phi_i}{\phi_{i'}}} L_{ij} \sqrt{\frac{\phi_j}{\phi_{j'}}}.
\label{eq:Lprimesym}
\end{equation}
Note that $L'$ remains symmetric if the original Laplacian $L$ is symmetric.

If asymmetric basis vectors are used by rescaling $\langle i'|\to w_{i'} \langle i'|$ and $|i'\rangle \to w_{i'}^{-1} |i'\rangle$ in Eq. (5), the resulting renormalized Laplacian $L'_{\rm A}$ differs in form from the one $L'_{\rm S}$ obtained using symmetric basis vectors. However, they are related by a similarity transformation: $L'_{\rm A} = W L'_{\rm S} W^{-1}$, and thus share the same eigenvalues. Consequently quantities such as the trace of propagators $Z^{(\ell)}(t)$, shown in Fig. 1(a), are invariant under the choice of normalization. 
Likewise, the adjacency matrix given by $A'_{i'j'} \equiv -N' \langle 0|i'\rangle \langle i'|L|j'\rangle \langle j'|0\rangle$  is also independent of the normalization, since the left and right basis vector scalings cancel out.  Therefore the structural features of the coarse-grained hypergraphs shown in Figs.~2 and 3 are unaffected by the normalization scheme. The solutions to the renormalized diffusion equation takes the form $x'_{i'}(t') = \sum_{j'} \langle i'|\rho'(t') |j'\rangle \langle j'|x(0)\rangle$,
where $\rho'(t') = e^{-L't'} = \sum_{\lambda'} e^{-\lambda' t'} |\lambda'\rangle \langle \lambda'|$. While this solution depends on the normalization of basis vectors, the relative error  compared with the coarse-grained original solution, defined as 
\begin{equation}
    \delta x_{i'}(t) \equiv \frac{x'_{i'}(t') - x_{i'}}{x_{i'}(t)} = \frac{\sum_{j'} \langle i'|\rho'(t') |j'\rangle \langle j'|x(0)\rangle - \sum_{i\in i',j}\langle i'|i\rangle \langle i|\rho(t)|j\rangle \langle j|x(0)\rangle}{\sum_{i\in i',j}\langle i'|i\rangle \langle i|\rho(t)|j\rangle \langle j|x(0)\rangle},
\end{equation}
is invariant against the choice of normalization; $|j'\rangle\langle j'|$ is unaffected by rescaling and any normalization dependence of $\langle i'|$ cancels out between the numerator and denominator.  Finally, the identification of local-equilibrium pairs, based on the propagator ratio in Eq. (9), is also unaffected by basis normalization. For the renormalized Laplacian, the ratio is given by 
\begin{equation}
   \frac{\rho'_{i'j'}(t'_c)\rho'_{j'i'}(t'_c)}{\rho'_{i'i'}(t'_c)\rho'_{j'j'}(t'_c)} =
    \frac{\langle i'|\rho'(t'_c)|j'\rangle \langle j'|\rho'(t'_c)|i'\rangle}{\langle i'|\rho'(t'_c)|i'\rangle \langle i'|\rho'(t'_c)|i'\rangle}
\end{equation}
which involves both left and right basis vectors and thus remains unchanged under normalization rescaling. 
 
\subsection{Timescale $t_c$ as a resolution parameter: Information entropy}
\label{sec:entropy}

To determine the coarse-graining timescale $t_c$, we utilize the information entropy $S(t)$. As we will show in this section, $S(t)$ represents the number of relevant (non-negligible) eigenmodes at time $t$.
We define $t_c$  as the time when $S(t)$ reaches a predefined threshold allowing  us  to identify and connect local-equilibrium pairs of nodes and obtain a target number $N'=(1-f)N$ of connected components (supernodes)  with $f$ a constant representing the reduction fraction.

\begin{figure}[h]
    \centering \vskip 2mm
    \includegraphics[width=0.6\textwidth]{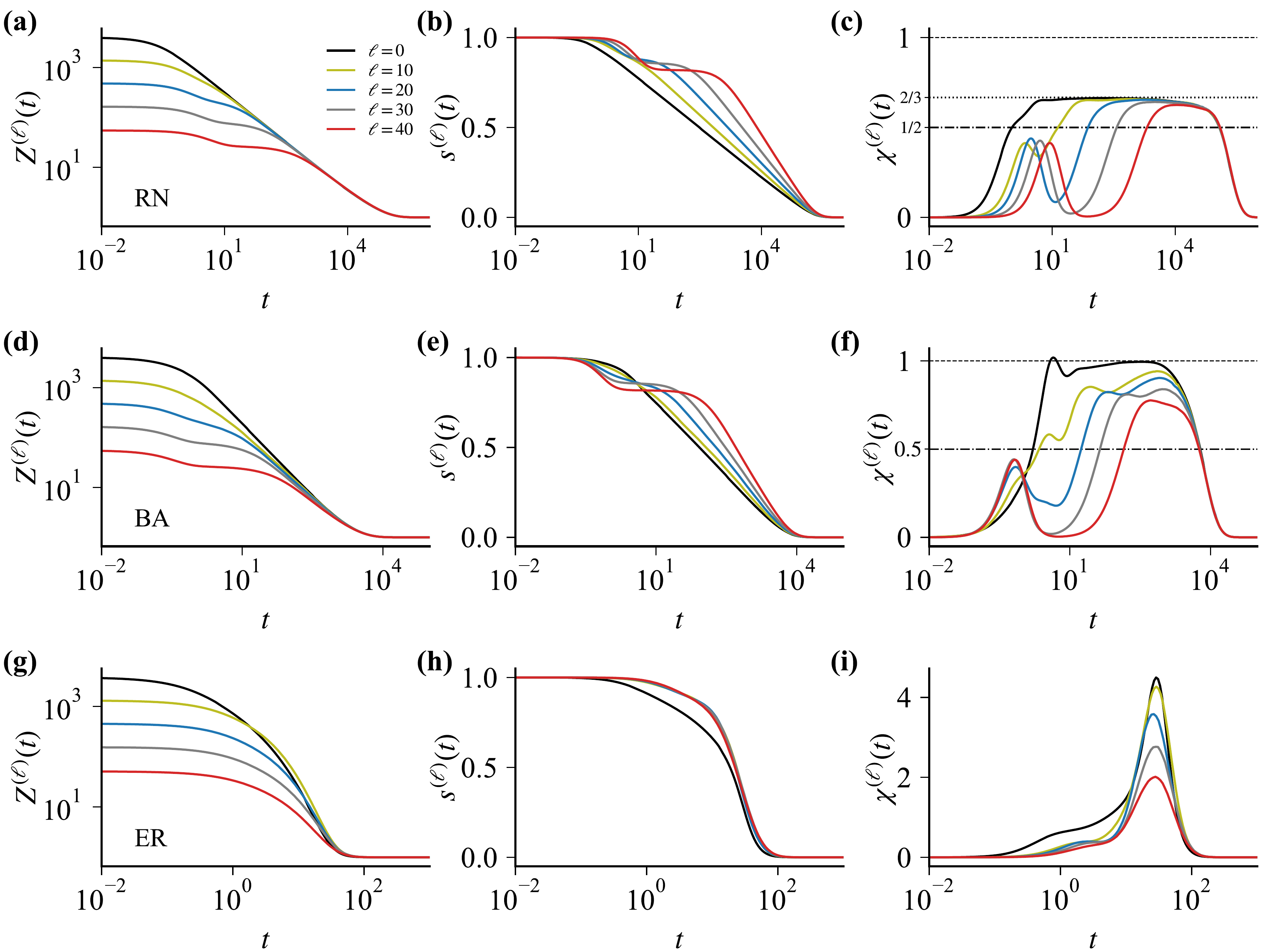}
    \caption{The trace of diffusion propagator $Z^{(\ell)}(t)$, the normalized information entropy $s(t)=\frac{S(t)}{\ln N}$, and its time derivative (heat capacity) $\chi(t) = -\frac{dS}{d\ln t}$ at several RG steps $\ell$ (with $f=0.1$)  for (a-c) RN hypertrees, (d-f) BA hypertrees, and (g-i) the giant components of  ER hypergraphs, initially of $V=2000$ vertices and $E=2000$ hyperedges. See the main text and Secs.~\ref{sec:entropy} and~\ref{sec:model} for their  details.  The mean connectivity  is $\bar{q}\approx 2$ in (a--f) and $\bar{q}=3$ in (g--i). The results were averaged over $1000$ hypergraph realizations. 
    }
    \label{fig:SC}
\end{figure}

The information entropy $S(t)$ is defined in terms of the normalized propagator $\tilde{\rho}(t) \equiv \frac{\rho(t)}{Z(t)}$, where the normalization factor is the trace of the unnormalized propagator  $Z(t)\equiv \mathrm{Tr} \rho(t)$ as introduced in Eq.~\eqref{eq:Ztdef} as
\begin{equation}
    S(t) \equiv -\langle \ln \tilde{\rho}(t)\rangle_t = \ln Z(t) + t\langle \lambda\rangle_t,
    \label{eq:S}
\end{equation}
in which $\langle A\rangle_t \equiv \mathrm{Tr} [A \tilde{\rho}(t)]$. Evaluating $Z(t) = \sum_n e^{-\lambda_n t}$ with $\lambda_1\leq \lambda_2\leq \cdots \leq \lambda_N$ the eigenvalues of the Laplacian sorted in the ascending order, one finds the sum  dominated by the contributions from the eigenvalues around the average $\langle \lambda\rangle_t$. This motivates the definition of the effective number of relevant modes,  $N_{\rm eff}(t)$, through the relation \begin{equation}
    Z(t)= N_{\rm eff}(t) e^{-t \langle \lambda\rangle_t}.
\end{equation}
Comparing  with Eq.~\eqref{eq:S}, we find 
\begin{equation}
    S(t)=\ln N_{\rm eff}(t).
    \label{eq:Neff}
\end{equation}
Thus $Z(t)$ and $N_{\rm eff}(t)$ decreases from $N$ to $1$ as $t$ increases, reflecting the diminishing number of relevant modes over time. See the behaviors of $Z(t)$ in Fig.~\ref{fig:SC} (a,d,g).

In this study, for a given reduction fraction $f$, we define $t_c$ as the time when the entropy $S(t)$ satisfies 
\begin{equation}
    S(t_c) = \ln [(1-\kappa f)N].
\end{equation}
Here  $\kappa$ is a constant that ensures a sufficient degree of local equilibrium and set to $\kappa=4$ throughout the present study except for Fig.~\ref{fig:kappa1} where the results obtained with $\kappa=1$ are presented for comparison. This means that the effective number of relevant eigenmodes is $N_{\rm eff}(t_c)=(1-\kappa f)N$. Then we connect node pairs $(i,j)$ sequentially in the descending order of the propagator ratio 
\begin{equation}
    \frac{\rho_{ij}(t_c) \rho_{ji}(t_c)}{\rho_{ii}(t_c)\rho_{jj}(t_c)}.
    \label{eq:ratio_localeq}
\end{equation}
until the total number of connected components is reduced to $N' = (1-f)N$. 
This procedure is based on the assumption that node pairs with the highest propagator ratio are already in local equilibrium at  $t_c$, and choosing $\kappa\geq 1$ prevents premature merging of unequilibrated pairs.

In Euclidean lattices and scale-invariant networks~\cite{Villegas:2023aa,poggialini2024}, the spectral density function $p(\lambda)$ defined in Eq.~\eqref{eq:pdef} exhibits a power-law form:
\begin{equation}
    p(\lambda) \sim \lambda^{\frac{d_{\rm s}}{2}-1}
    \label{eq:ds}
\end{equation}  
over a wide range of $\lambda$, i.e., $\lambda_2\leq \lambda < \lambda_*$, with $d_{\rm s}$ called the spectral dimension. This leads to
\begin{equation}
    Z(t) = \sum_n e^{-\lambda_n t} = N\int d\lambda p(\lambda) e^{-\lambda t} \sim N t^{-d_{\rm s}/2},
\end{equation}
and 
\begin{equation}
    \langle \lambda\rangle_t = \frac{\int d\lambda p(\lambda) \lambda e^{-\lambda t}}{\int d\lambda p(\lambda) e^{-\lambda t}} = \frac{d_{\rm s}}{2} t^{-1}.
\end{equation}
See Fig.~\ref{fig:SC} (a,d). Substituting into Eqs.~\eqref{eq:S} and \eqref{eq:Neff}, we obtain
\begin{equation}
    S(t) \simeq \ln N -\frac{d_{\rm s}}{2} \ln t + {\rm constant},
    \label{eq:Sds}
\end{equation}
and 
\begin{equation}
    N_{\rm eff}(t)\simeq N t^{-\frac{d_{\rm s}}{2}},
    \label{eq:Neffdecay}
\end{equation}
which hold over a broad  range of time $t_* < t<t_{\rm eq} = \lambda_2^{-1}$ [Fig.~\ref{fig:SC} (b) and (e)]. In this time regime $t_*<t<t_{\rm eq}$, the time derivative $\chi(t) \equiv -t\frac{dS}{dt}$ of the information entropy or the ``heat capacity" remains constant, that is, ``show a plateau''~\cite{Villegas:2023aa,poggialini2024}
\begin{equation}
    \chi \simeq \frac{d_{\rm s}}{2}
    \label{eq:chiconst}
\end{equation}
as shown in Fig~\ref{fig:SC}(c,f). In some networks and hypergraphs, Eq.~\eqref{eq:Sds} or \eqref{eq:chiconst} holds only for a narrow range of time, preventing a well-defined spectral dimension [Fig.~\ref{fig:SC}(h,i)].

\section{Models for hypergraphs}
\label{sec:model}

In this section we describe the procedures to construct model hypergraphs to which we apply the EqLRG. Before presenting them, we first introduce the bipartite-network representation of hypergraphs. Let us consider a hypergraph, consisting of $V$ vertices and $E$ hyperedges with an incidence matrix $B_{ve}$, representing whether a vertex $v$ belongs to a hyperedge $e$ or not by $B_{ve}=1$ or $0$, respectively. This hypergraph can be represented by a bipartite network of $N=V+E$ nodes, including $V$ vertex-type nodes and $E$ hyperedge-type nodes that are connected by $M = \sum_{v,e} B_{ve}$ links. Its adjacency matrix $A_{ij}$ is of dimension $N\times N$ with  $A_{ij}=1$ if $B_{ve}=1$ with $\{i,j\}=\{v,e\}$ and $0$ otherwise. The mean connectivity is evaluated as $\bar{q} = N^{-1}\sum_{ij} A_{ij} = \frac{2M}{N}$. For the hypergraph, the mean vertex degree  is given by $\bar{k} = V^{-1}\sum_{v} q_v = V^{-1}\sum_{v,e} B_{ve} = \frac{M}{V}$ and the mean hyperedge cardinality is $\bar{c} = E^{-1}\sum_e c_e = E^{-1}\sum_{e,v} B_{ve}=\frac{M}{E}$. 

\subsection{Random (RN) hypertrees}
We construct a random (RN) hypertree using the following procedure, slightly modifying~\cite{chen2000randomtree}: 
\begin{enumerate}[label=(\roman*)]
\item Nodes, that will become either vertices or hyperedges, are numbered by $i=1,2,\ldots, N=V+E$. A Pr\"{u}fer sequence of length $N-2$ is generated by randomly selecting $N-2$ node numbers with replacement. 
\item Stubs (half-links) are assigned to each node $i$,  as many as the number of times the number $i$ appears in the Pr\"{u}fer sequence, plus one. This results in a total  of $2N-2$ stubs that will be paired to form $N-1$ links at the next steps. 
\item The lowest-numbered node among the nodes having just one stub left for pairing and the first node, say $j$, in the Pr\"{u}fer sequence are connected.  Both nodes are deprived of one stub, respectively, and the first element $j$ is deleted from the Pr\"{u}fer sequence. 
\item Step (iii) is repeated $N-2$ times until the Pr\"{u}fer sequence is emptied. Then two nodes with one stub each remain. By connecting these two nodes, the construction of a spanning tree with $N$ nodes and $N - 1$ links is completed.
\item A randomly-selected node is classified as a vertex, and its nearest-neighbor nodes become hyperedges. Their nearest-neighbors become vertices, and so on, alternating labels recursively until all nodes are classified as either vertices or hyperedges.
\end{enumerate}

Since every node---whether a vertex or a hyperedge---is selected $N-2$ times,  each with probability $N^{-1}$ in the construction of the Pr\"{u}fer sequence, and is additionally assigned one stub, the resulting degree $k$ of vertices and cardinality $c$ of hyperedges minus one follow Poisson distributions with mean $(N-2)\frac{1}{N}\simeq 1$ for large $N$ given by
$P(k) = \frac{e^{-1}}{(k-1)!}$ and $P(c) = \frac{e^{-1}}{(c-1)!}$ respectively. These distributions yield  the  expected mean  values of $\langle \bar{k}\rangle= \sum_{k=1}^\infty k P(k) = 2$ and $\langle\bar{c}\rangle= \sum_{c=1}^\infty c P(c)=2$, which is consistent with the expected average degree and cardinality  in tree structures. The variances are  $\bar{k^2} - \bar{k}^2 = 1$ and $\bar{c^2} - \bar{c}^2 = 1$, leading to heterogeneity indices $\eta_k = \frac{\sqrt{\bar{k^2}-\bar{k}^2}}{\bar{k}} = \frac{1}{2}$ and $\eta_c = \frac{\sqrt{\bar{c^2}-\bar{c}^2}}{\bar{k}} = \frac{1}{2}$. The heat capacity of these RN hypertrees displays a plateau at $\chi=\frac{2}{3}$ [Fig.~\ref{fig:SC}(c)].

\subsection{Barab\'{a}si-Albert-type (BA) hypertrees}
We use the following procedures to construct a scale-free hypertree with equal numbers of vertices and hyperedges ($V=E$) and power-law degree and cardinality distributions, the ensemble of which we call BA hypertrees in the main text:
\begin{enumerate}[label=(\roman*)]
\item We begin with a small bipartite network of one vertex-type node and one hyperedge-type node. While the final configuration may be influenced by this initial setup, such dependence diminishes as the network size grows sufficiently large. 
\item We alternately add either a vertex or an hyperedge to the bipartite network. Each newly added node is connected to $m$ existing nodes of the opposite type, where the selection is made based on a probability proportional to the degree of the existing nodes. Here we set $m=1$ to construct a bipartite tree.
\item This iterative process continues until the desired number of vertices and edges is reached. 
\end{enumerate}

In the present study we simply adopt the symmetric scenario that the number of vertices and the number of hyperedges are identical, as in the RN hypertrees,  so both the mean vertex degree $\bar{k}$ and the mean hyperedge cardinality $\bar{c}$ are $2$. Significantly, the degree distribution of the vertices and the cardinality distribution of the hyperedges in this hypergraph both adhere to a power-law distribution, with an exponent of $3$, i.e., $P(k)\sim k^{-3}$ and $P(c)\sim c^{-3}$, mirroring the characteristics of the original BA network. Their variances diverge with $V$ and $E$, respectively, as $\bar{k^2}-\bar{k}^2\sim \ln V$ and $\bar{c^2}-\bar{c}^2\sim \ln E$. The spectral dimension of the BA hypertrees is $d_s=2$ as seen from the plateau of the heat capacity shown in Fig.~\ref{fig:SC}(f) (black curve).

The methodology outlined can be extended to generate hypergraphs with differing numbers of vertices and edges, thus allowing for asymmetry in their structures. A viable approach to introduce asymmetry involves introducing the probability of alternately adding vertices and edges, or alternatively, implementing simultaneous multiple additions of each type during the construction process.

\subsection{Static-model (ST) and Erd\"{o}s-R\'{e}nyi (ER) hypergraphs} 
By generalizing the Erd\H{o}s-R\'{e}nyi graphs, Goh {\it et al.} introduced the so-called static-model for random scale-free networks~\cite{Goh2001_Static}. Extending the model to hypergraphs, we use the following procedure to construct a static-model (ST) hypergraph with a power-law or Poisson degree and cardinality distributions: 
\begin{enumerate}[label=(\roman*)]
\item We begin with $V$ vertices and $E$ hyperedges, both isolated. 
\item Fitness values $f_v$ and $f_e$ are assigned to vertices and hyperedges numbered by $v=1,2,\ldots, V$ and $e=1,2,\ldots, E$, respectively, as
$f_v = \frac{v^{-\alpha_{\rm V}}}{\sum_{n=1}^{V}n^{-\alpha_{\rm V}}}$ and $f_e = \frac{e^{-\alpha_{\rm E}}}{\sum_{n=1}^{E}n^{-\alpha_{\rm E}}}$ with $0\leq \alpha_{\rm V}, \alpha_{\rm E}<1$.
\item A pair of vertex $v$ and hyperedge $e$ are randomly selected with probability $f_v f_e$ and connected by a link if they are not already connected.
\item This process is repeated $M$ times. 
\end{enumerate}

Consequently $M$ vertex-hyperedge pairs are connected, leading to $\bar{k}=\frac{M}{V}$ and $\bar{c}=\frac{M}{E}$. The resultant vertex degree and hyperedge cardinality distributions are given by 
$P(k)\sim k^{-\gamma_{\rm V}}$ and $P(c)\sim c^{-\gamma_{\rm E}}$ with $\gamma_{\rm V} = 1+ \frac{1}{\alpha_{\rm V}}$ and $\gamma_{\rm E} = 1+\frac{1}{\alpha_{\rm E}}$ if both $\alpha_{\rm V}$ and $\alpha_{\rm E}$ are positive. In the main text, we take $\alpha_{\rm V}=0$ and $\alpha_{\rm E}=0$ to select each vertex-hyperedge pair with uniform probability and thereby construct the Erd\"{o}s-R\'{e}nyi (ER) hypergraphs with $M=3V=3E$ or equivalently $\bar{k}=\bar{c}=3$. These ER hypergraphs have Poisson degree and cardinality distributions $P(k) = \frac{\bar{k}^k e^{-\bar{k}}}{k!}$ and $P(c) = \frac{\bar{c}^c e^{-\bar{c}}}{c!}$, which have variances $\bar{k^2}-\bar{k}^2 = \bar{k}=3$ and $\bar{c^2}-\bar{c}^2 = \bar{c}=3$, resulting in heterogeneity indices $\eta_k=\eta_c=\frac{1}{\sqrt{3}}$.

Also in Fig.~\ref{fig:SThyper}, we present the results obtained for ST hypergraphs with $\alpha_{\rm V}=\frac{2}{3}$, i.e., $\gamma_{\rm V}=\frac{5}{2}$ and $\alpha_{\rm E}=0$, for which the degree distribution takes a power-law $P(k)\sim k^{-5/2}$ and the cardinality distribution takes the Poissonian form $P(c) = \frac{\bar{c}^c e^{-\bar{c}}}{c!}$ with $\bar{k}=\bar{c}=3$. Therefore the variance of the vertex degree diverges while that of the hyperedge cardinality is finite. These hypergraphs do not have a well-defined spectral dimension; The heat capacity do not show a wide plateau in Fig.~\ref{fig:SC}(i) or ~\ref{fig:SThyper}(b).

\section{Applications of EqLRG}
\label{sec:app_hypergraph}

In this section we present results from applying the EqLRG to various model and real-world hypergraphs. Through recursive applications,  we obtain the renormalized Laplacians $L'=L^{(\ell+1)}$ from $L=L^{(\ell)}$ by Eq. (5) of the main text across successive RG steps $\ell$. These results illustrate the coarse-graining process and  demonstrate  the robustness of our findings presented in the main text.

\subsection{A small toy hypergraph}

\begin{figure}[h]
    \centering
    \includegraphics[width=0.6\textwidth]{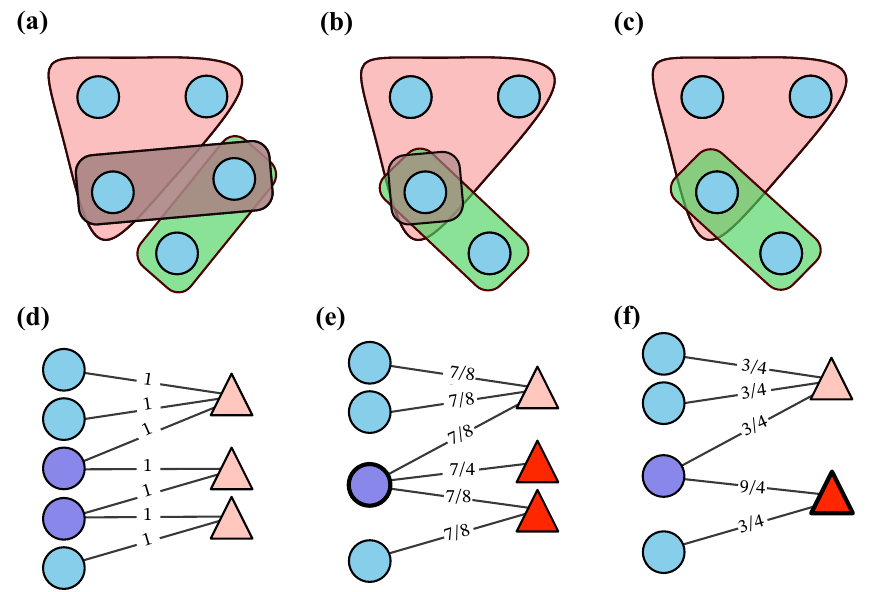} 
    \caption{Illustration of EqLRG for a hypergraph with $5$ vertices and $3$ hyperedges. Two vertices in (a) are merged into a single supervertex at the first RG step in (b), followed by the merging of two hyperedges  into a superhyperedge at the second step in (c), via recursive application of the EqLRG with $f=0.1$. Panels (d,e,f) show the corresponding bipartite representations of (a,b,c), respectively, with original and merged vertices (hyperedges) colored violet (red). In (d), all connected vertex-hyperedge pairs have unit weight ($B_{ve}=1$).  After the first coarse-graining in (e), the hyperedge originally connected to both violet vertices is linked to the supervertex with weight $\frac{7}{8}(1+1)=\frac{7}{4}$ while all other connections have weight $\frac{7}{8}$. After the second coarse-graining in (f), the two red hyperedges are merged, yielding a new connection weight with the supervertex of $\frac{6}{7}(\frac{7}{4}+\frac{7}{8})=\frac{9}{4}$.
    }
    \label{fig:illust}
\end{figure}

To provide an intuitive understanding of the EqLRG, we illustrate its application to a small toy hypergraph in Fig.~\ref{fig:illust}. This example demonstrates how the EqLRG is performed on a bipartite-network representation of a hypergraph, highlighting the renormalization of the adjacency matrix. 

A hypergraph, shown in  Fig.~\ref{fig:illust}(a), can be represented as a bipartite network [Fig.~\ref{fig:illust}(c)]  consisting of $N=8$ nodes -- $V=5$ vertices and $E=3$ hyperedges. Each vertex-hyperedge connection is encoded in the incidence matrix with unit weight, i.e., $B_{ve}=1$ if $v$ and $e$ are connected and $B_{ve}=0$ otherwise.  Applying the EqLRG procedure with $f=\frac{1}{8}$, two hyperedges are merged into a superhyperedge, leading to a coarse-grained hypergraph in Fig.~\ref{fig:illust}(b). In its bipartite-network representation [Fig.~\ref{fig:illust}(d)] with $N'=7$ nodes, the renormalized incidence between the vertex $v$ originally connected to the two hyperedges and the new superhyperedge  $e'$ is given by $B'_{ve'}=\frac{7}{4}$, computed by Eq. (7) as the sum of  original weights $2$ modulated by the ratio $\frac{N'}{N}=\frac{7}{8}$. Other vertex-hyperedge pairs remain unchanged but their weights are changed to $\frac{7}{8}$ due to the modulation. 

\subsection{Accuracy in capturing the original diffusion dynamics}
\label{sec:accuracy}

To validate the renormalized diffusion given in Eq. (3) of the main text, we first compare the trace of the diffusion propagator, $Z^{(\ell)}(t) = \sum_{\lambda^{(\ell)}} e^{-\lambda^{(\ell)} t^{(\ell)}}$, where $\lambda^{(\ell)}$'s are the eigenvalues of the Laplacian matrix $L^{(\ell)}$ at RG step $\ell$, with the original one $Z(t) = Z^{(0)}(t)$ from the original dynamics in Eq. (1), which should closely match  in the late-time regime, provided that the small eigenvalues of the Laplacian are preserved. Figure 1(a) and Fig~\ref{fig:SC}(a,d,g) show that they indeed match well in the late-time regime. The rescaled time $t^{(\ell)}$ is related to the original time $t$ by $t^{(\ell)} = t \frac{\lambda_2^{(0)}}{\lambda^{(\ell)}_2}$ according to Eq. (6). 

\begin{figure}[h]
    \centering
    \includegraphics[width=0.5\textwidth]{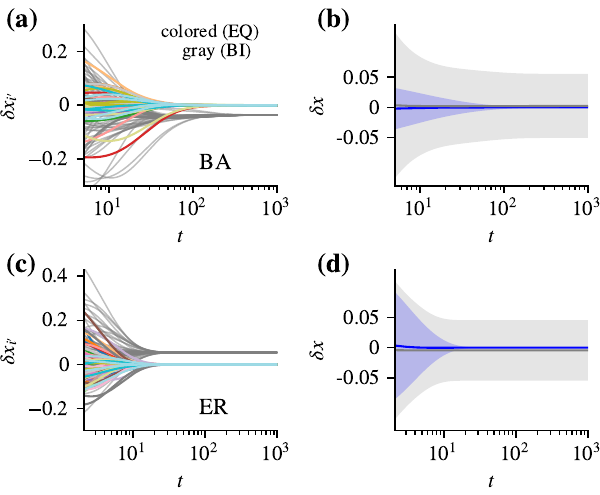}
    \caption{Relative error of the solutions ($x'_{i'}(t')$'s) to Eq. (3) compared to $x_i(t)$'s from Eq. (1) for (a,b) BA hypertrees and (c,d) ER hypergraphs, both of $V=100$ vertices and $E=100$ hyperedges. They are shown for both the binary (BI) and equilibrium-preserving (EQ) versions of the LRG with $f=0.1$. (a,c) Relative errors $\delta x_{i'}(t)$, as defined in the caption of Fig. 1, for all supernodes $i'$ in a single realization of the diffusion dynamics with a random initial condition. (b,d) The average (lines) and the standard deviation (shades) of the relative errors over all supernodes and multiple realizations, demonstrating higher accuracy of the EqLRG (blue) in preserving diffusion dynamics than the binary counterpart (gray). 
    }
    \label{fig:traj}
\end{figure}

More importantly, we find that the solutions $x'_{i'}(t')$'s of the renormalized diffusion equation [Eq. (3)] closely match the coarse-grained original solutions $x_{i'}(t)$'s (from Eq. (1)) as expected from Eq.~\eqref{eq:xprimex}.  To quantify their deviation, we define the {\it relative error} as
\begin{equation}
    \delta x_{i'}(t) \equiv \frac{x'_{i'}(t'(t)) - x_{i'}(t)}{x_{i'}(t)},
    \label{eq:deltax}
\end{equation}
where the rescaled time $t'$ and the original time $t$ are related by Eq. (6). As shown in Fig. 1(b) and Fig.~\ref{fig:traj}, the relative error decreases significantly  with time, confirming the validity of the renormalized diffusion under the EqLRG.

To examine whether such convergence of the two solutions depends on the choice of the quasi-complete basis and the renormalized Laplacian as given in Eq. (5), we consider an alternative LRG scheme - binary LRG. While supernodes are still constructed by connecting pairs with the highest ratio in Eq.~\eqref{eq:ratio_localeq} or Eq. (9),  the renormalized adjacency and Laplacian are defined as 
\begin{equation}
    A^{\prime {\rm (BI)}}_{i'j'} = \left\{
    \begin{array}{ll}
    1 & {\rm if} \ \sum_{i\in i', j\in j'}A_{ij}>0 \\
    0 & {\rm otherwise}
    \end{array}
    \right. 
    \ {\rm and} \ \ 
    L^{\prime {\rm (BI)}}_{i'j'} = q^{\prime {\rm (BI)}}_{i'} \delta_{i'j'} - A^{\prime {\rm (BI)}}_{i'j'},
    \label{eq:BI}
\end{equation}
where the connectivity of supernode $i'$ is defined as $q^{\prime {\rm (BI)}}_{i'} \equiv \sum_{j'} A^{\prime {\rm (BI)}}_{j'i'}$. The relative error $\delta x_{i'}^{\rm (BI)}(t)$ of the solutions to Eq. (3),  using $L^{\prime {\rm (BI)}}$, is computed via Eq.~\eqref{eq:deltax}. Since no definite basis exists for supernodes in this case, the coarse-grained original solutions are estimated by the average over constituent nodes as 
\begin{equation}
    x_{i'}(t) \equiv \frac{\sum_{i\in i'}x_i(t)}{\sum_{i\in i'}1}.
\end{equation}
As shown in Fig. 1(b) and Fig.~\ref{fig:traj}, the binary LRG yields noticeably larger relative errors compared to the EqLRG, indicating less accurate approximation of the original diffusion dynamics. 

\subsection{Supplementary EqLRG results for model hypergraphs}

Here we present supplementary EqLRG results for the model hypergraphs introduced in Sec.~\ref{sec:model}, complementing those in the main text. Additional analysis of the obtained results are also provided. 

\begin{figure}[h]
    \centering
    \includegraphics[width=0.65\textwidth]{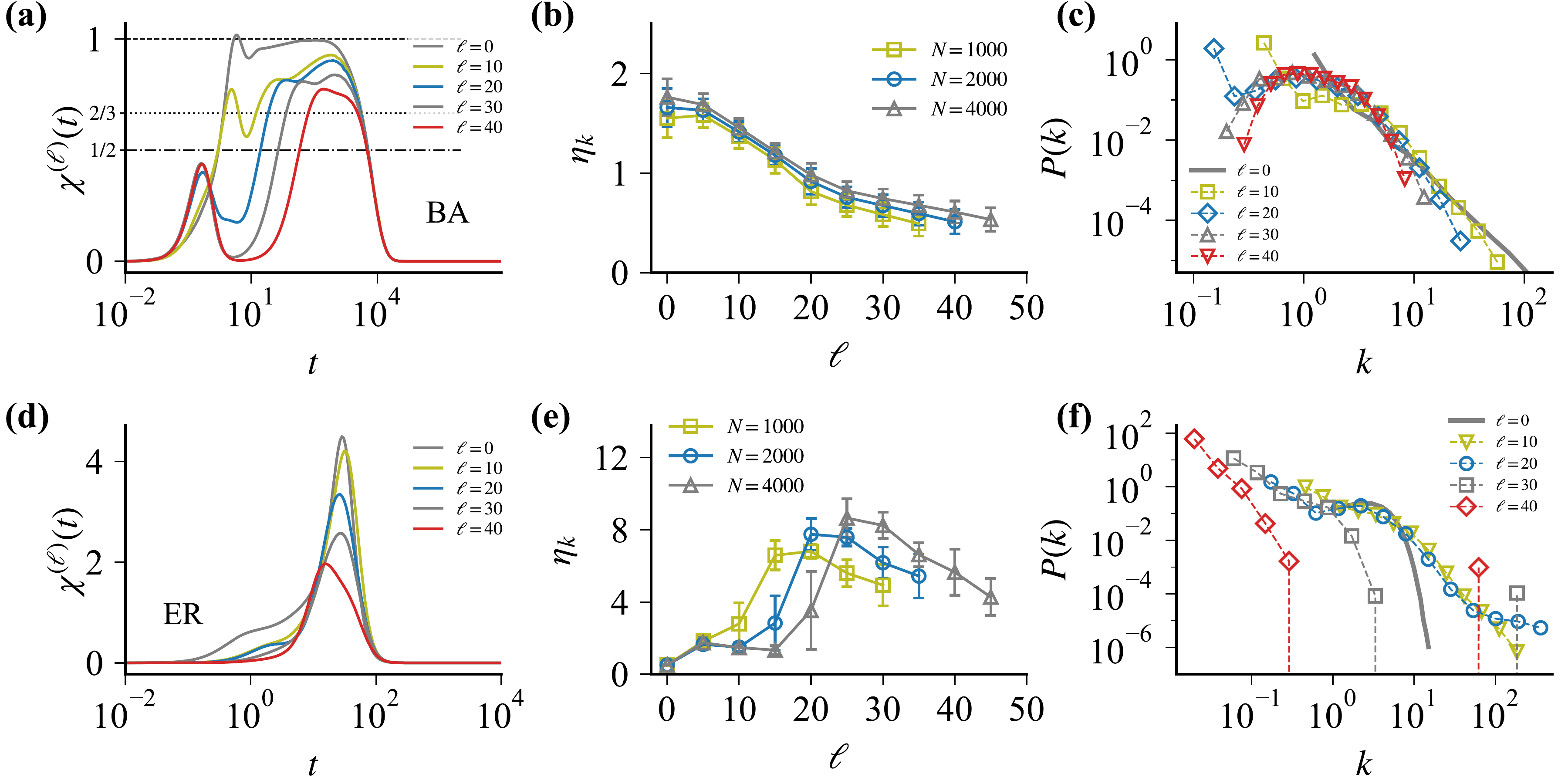} 
    \caption{Heat capcity $\chi(t)$, degree-heterogeneity index $\eta_k$, and degree distribution $P(k)$ at different EqLRG steps $\ell$ with $f=0.1$ and a smaller value $\kappa=1$ than the one used in the main text for (a-c) BA hypertrees and (d-f) ER hypergraphs with $V^{(0)}=E^{(0)}=2000$.
    }
    \label{fig:kappa1}
\end{figure}

$\bullet$ {\it EqLRG with different $\kappa$---}In the main text, we presented  the EqLRG results obtained with $f=0.1$ and $\kappa=4$ for the BA hypertrees and ER hypergraphs.  The results obtained with a smaller value of $\kappa=1$ shown in Fig.~\ref{fig:kappa1} remain qualitatively unchanged compared with the results obtained with $\kappa=4$ in the main text, supporting the robustness of our findings.

\begin{figure}[h]
    \centering
    \includegraphics[width=0.95\textwidth]{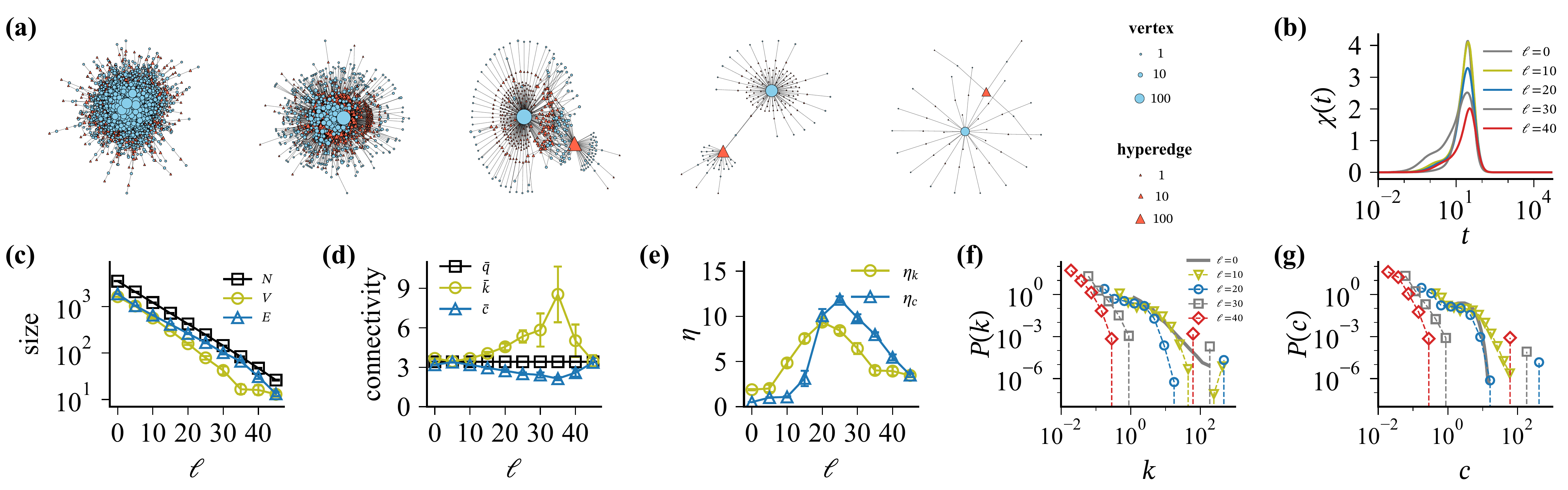} 
    \caption{Evolution of ST hypergraphs with $\gamma_{\rm V}=2.5$ and $\gamma_{\rm E}\to\infty$  under the EqLRG.    
     (a) Transformation of a ST hypergraph ($V= E = 2000$) under recursive EqLRG with $f = 0.1$ from the original one ($\ell = 0$) to the transformed ones at RG steps $\ell = 10,20,30$ and $40$ (from left to right). They are represented as bipartite networks, where circles and triangles represent vertices and hyperedges, with sizes increasing with degree and cardinality, respectively. (b)  Heat capacity $\chi(t)$ as a function of time $t$ at different RG steps $\ell$. 
     (c) Number of supernodes ($N$), supervertices ($V$), and superhyperedges ($E$) versus RG steps $\ell$. 
     (d) Mean connectivity ($\bar{q}$), mean vertex degree ($\bar{k}$), and mean hyperedge cardinality ($\bar{c}$).  
     (e) Heterogeneity indices $\eta_k$ and $\eta_c$. 
     (f) Degree distribution $P(k)$ and (g) Cardinality distribution $P(c)$ at various RG steps $\ell$. 
}
    \label{fig:SThyper}
\end{figure}

$\bullet$ {\it EqLRG for ST hypergraphs---}Note that in Fig. 3 of the main text, we presented EqLRG results only for vertex degree, i.e., the degree heterogeneity $\eta_k$ and the degree distribution $P(k)$ since the statistics of degree $k$ and cardinality $c$ are identical in RN and BA hypertrees and ER hypergraphs we considered. In Fig.~\ref{fig:SThyper}, we present the EqLRG results for ST hypergraphs with different initial distributions for $k$ and $c$: namely, a power-law degree distribution $P(k)\sim k^{-2.5}$ and the Poissonian cardinality distribution $P(c) = \frac{\bar{c}^c e^{-\bar{c}}}{c!}$. As the RG proceeds, vertices of larger degrees and hyperedges of larger cardinalities emerge, eventually forming an ultra-hub vertex and a ultra-hub hyperedge at late RG steps [Fig.~\ref{fig:SThyper}(a)]. The number of supervertices  decreases faster than that of superhyuperedges, leading to $\bar{k}>\bar{c}$ [Fig.~\ref{fig:SThyper}(c,d)]. This occurs because hub vertices are more abundant than hyperedges of large cardinality initially and they merge with many vertices in the neighborhood, forming larger hub supervertices through coarse-graining. The heterogeneity indices $\eta_k$ and $\eta_c$ increase as the RG progresses while they decline at later RG steps due likely to the significant reduction in the number of supernodes, downscaling the adjacency matrix via Eq. (7).  Accordingly, both the degree and cardinality distributions broaden---especially cardinality distribution---and develop humps [Fig.~\ref{fig:SThyper}(f,g)].  These humps are related to the ultra-hub vertex and hyperedge observed in Fig.~\ref{fig:SThyper}(a), implying a condensaation of connections.

\subsection{Cluster-size and connectivity of supernodes}
In the EqLRG applied to a hypergraph, all pairs of vertices or of hyperedges that are in {\it local-equilibrium}, i.e., those displaying the largest values of Eq. (9) of the main text, are first identified.  Each supervertex or superhyperedge corresponds to a connected component of these  local-equilibrium pairs of vertices (hyperedges), where any two such pairs are connected by a path. As the diffusion timesclae $t_c$ increases, more pairs enter local equilibrium, leading to the aggregation of these components. 

\begin{figure}[h]
    \centering \vskip 2mm
    \includegraphics[width=0.45\textwidth]{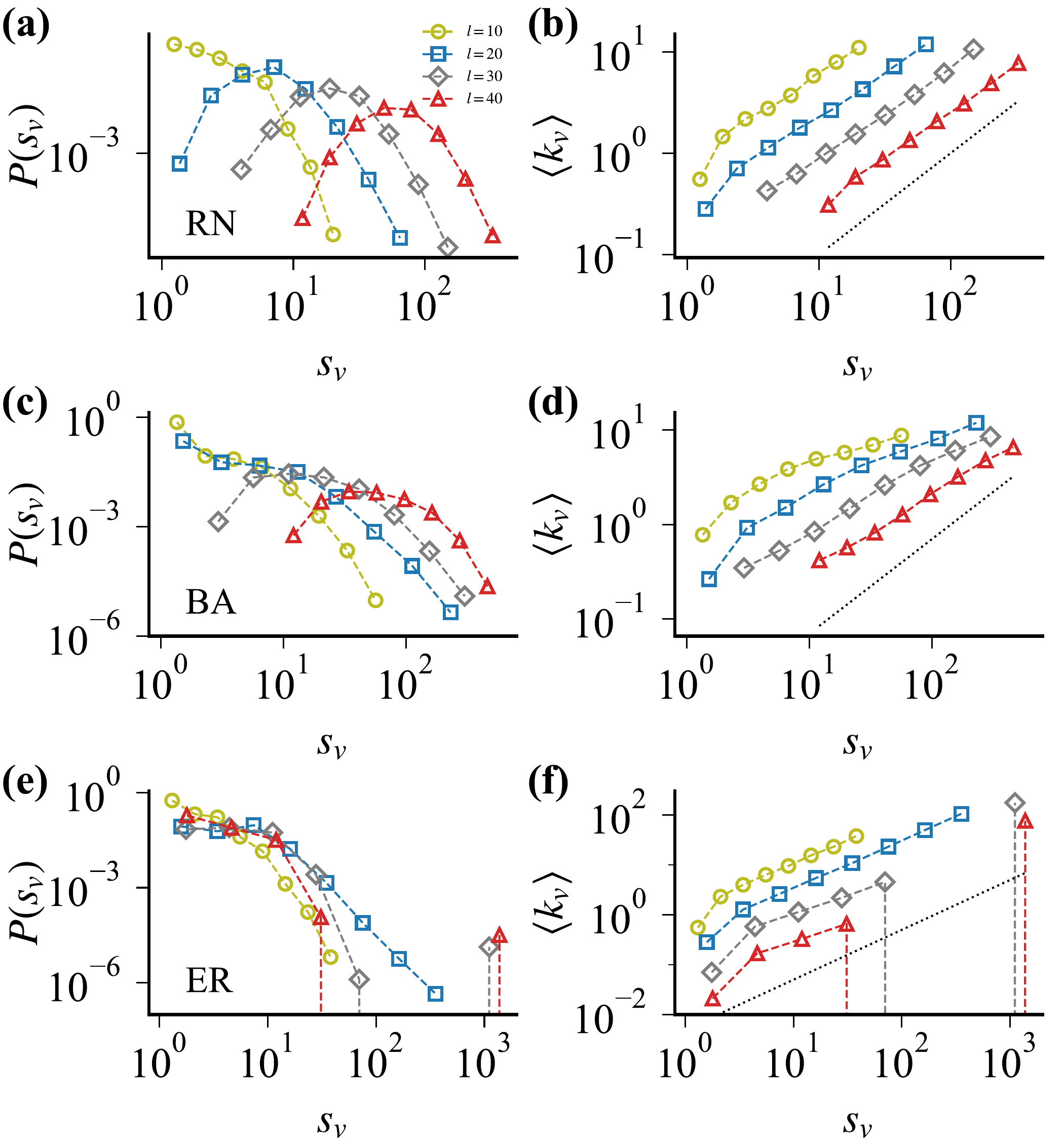} 
    \caption{    
    Cluster-size and connectivity of supernodes under the EqLRG. Each row present the cluster-size distribution and the relation between degree and cluster-size during the EqLRG for (a,b) RN hypertrees, (c,d) BA hypergraphs, and (e,f) ER hypergraphs. (a,c,e) Distribution of the cluster-size of supervertex $s_v$, (b,d,f) degree versus cluster-size of supervertex. The dotted lines have slope $1$.
     Each plot is obtained with $1000$ realization of size $V=E=2000$ through EqLRG with $f=0.1$. 
     }
    \label{fig:clustersize}
\end{figure}

We define the {\it cluster size} $s_v$ ($s_e$) of a supervertex $v$ (a superhyperedge $e$) as the number of nodes it contains. The distribution $P(s_v)$ (and $P(s_e)$ as well) remains narrow in RN hypertrees and evolves toward narrower forms in BA hypertrees  over increasing RG steps $\ell$ [Fig.~\ref{fig:clustersize} (a,c)]. For ER hypergraphs, the cluster-size distributions first evolve toward power-law forms and then develop humps at later RG stages [Fig.~\ref{fig:clustersize} (e)]. In all these hypergraphs, the degree or cardinality of a supernode scales approximately linearly with its cluster size [Fig.~\ref{fig:clustersize}(b,d,f)], suggesting that the broadening of the renormalized degree and cardinality distributions originates from the cluster-size heterogeneity. 

\subsection{Binary LRG}
\label{sec:bilrg}

\begin{figure}[h]
    \centering
    \includegraphics[width=0.8\textwidth]{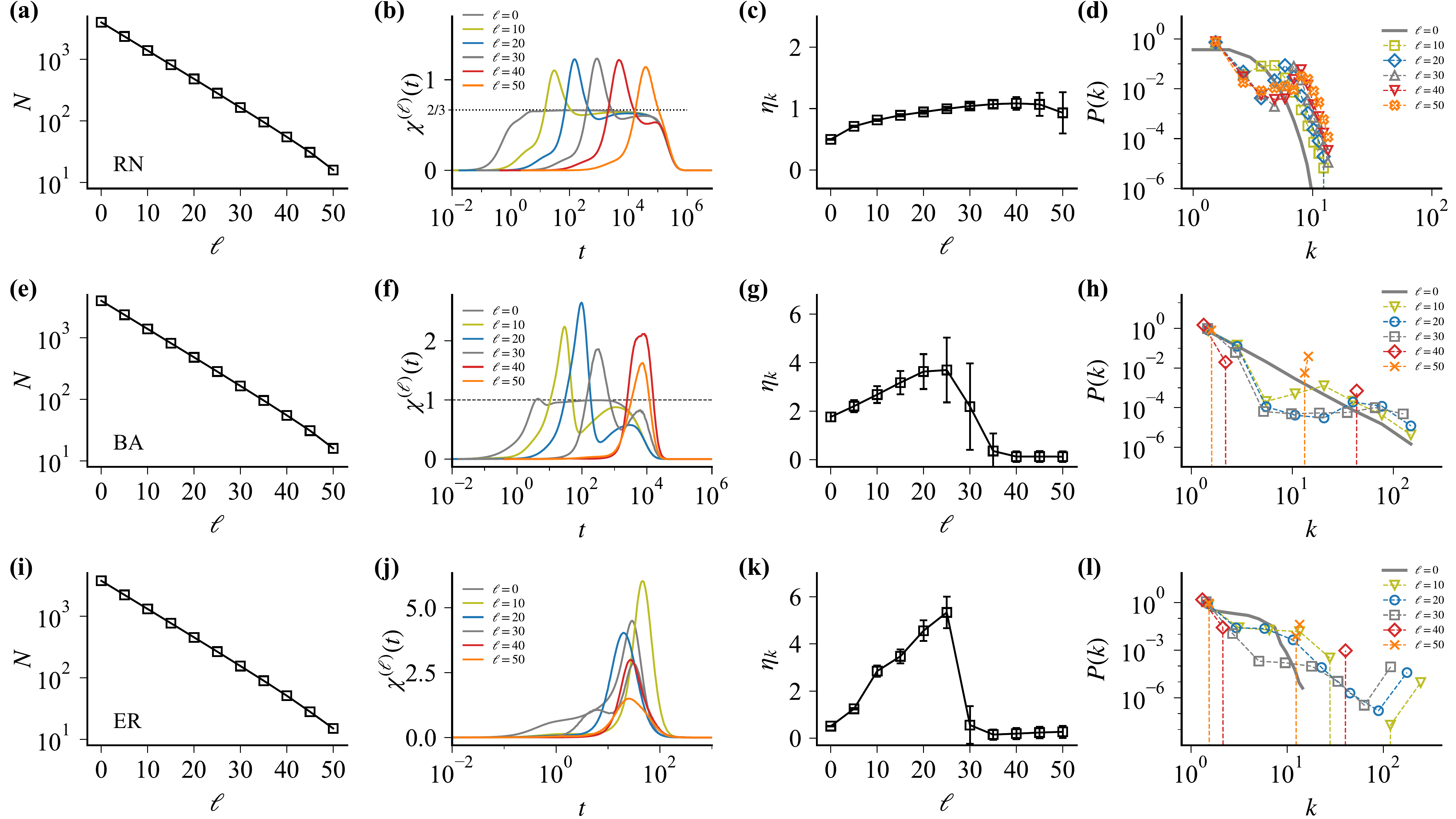} 
    \caption{
    Evolution of hypergraphs under the binary LRG with $f=0.1$. The results for (a-d) RN hypertrees, (e-h) BA hypertrees, and (i-l) ER hypergraphs are shown.
    (a,e,i) Number of supernodes ($N$) versus RG steps $\ell$.
    (b,f,j) Heat capacity $\chi(t)$ versus $t$ for different $\ell$’s. (c,g,k) Degree heterogeneity index $\eta_k$ versus RG steps $\ell$. (d,h,l) Degree distribution $P(k)$ at different RG steps $\ell$.
    }
    \label{fig:binaryLRG}
\end{figure}

The transformation of  hypergraphs under the binarly LRG  defined in Eq.~\eqref{eq:BI} is shown in Fig.~\ref{fig:binaryLRG}, revealing significant  differences from the results of the EqLRG. 

Most remarkably, while the effective spectral dimensions of RN and BA hypertrees converge to $d_s\approx \frac{4}{3}$ under the EqLRG---as indicated by the plateau at $\chi\approx \frac{2}{3}$ observed across RG steps shown in Fig. 2(e) and ~\ref{fig:SC}(c,f)---this plateau is absent under the binary LRG. In the binary scheme,  the $2/3$-plateau persists only in RN hypertrees[Fig.~\ref{fig:binaryLRG}(b)],  whereas BA hypertrees fail to exhibit a plateau in $\chi(t)$ as RG progresses [Fig.~\ref{fig:binaryLRG}(f)]. Additionally, both RN and BA hypertrees shown an early-time peak in $\chi(t)$ that exceeds the corresponding peaks in the EqLRG (which remains below $\chi=\frac{1}{2}$). For ER hypergraphs, $\chi(t)$ has initially a peak, which first  increases and then decreases later under the binary LRG [Fig.~\ref{fig:binaryLRG}(j)], contrasting with its steady decrease under the EqLRG.

Beyond spectral properties, the evolution of connectivity structure also differs substantially  between the two LRG schemes. Under the binary LRG, the degree heterogeneity index $\eta_k$ increases over a broad range of RG steps for all the considered cases [Fig.~\ref{fig:binaryLRG}(c,g,k)] in contrast to the EqLRG, where  $\eta_k$ remains low or decreases in RN and BA hypertrees and increases only for ER hypergraphs. Accordingly, the degree distribution (as well as the cardinality distribution) broadens with RG steps under the binary LRG[Fig.~\ref{fig:binaryLRG}(d,h,l)], whereas under the EqLRG, it broadens only for ER or ST hypergraphs and converges to Poissonian forms for RN and BA hypertrees with low spectral dimensions.

\begin{figure}[h]
    \centering
    \includegraphics[width=0.45\textwidth]{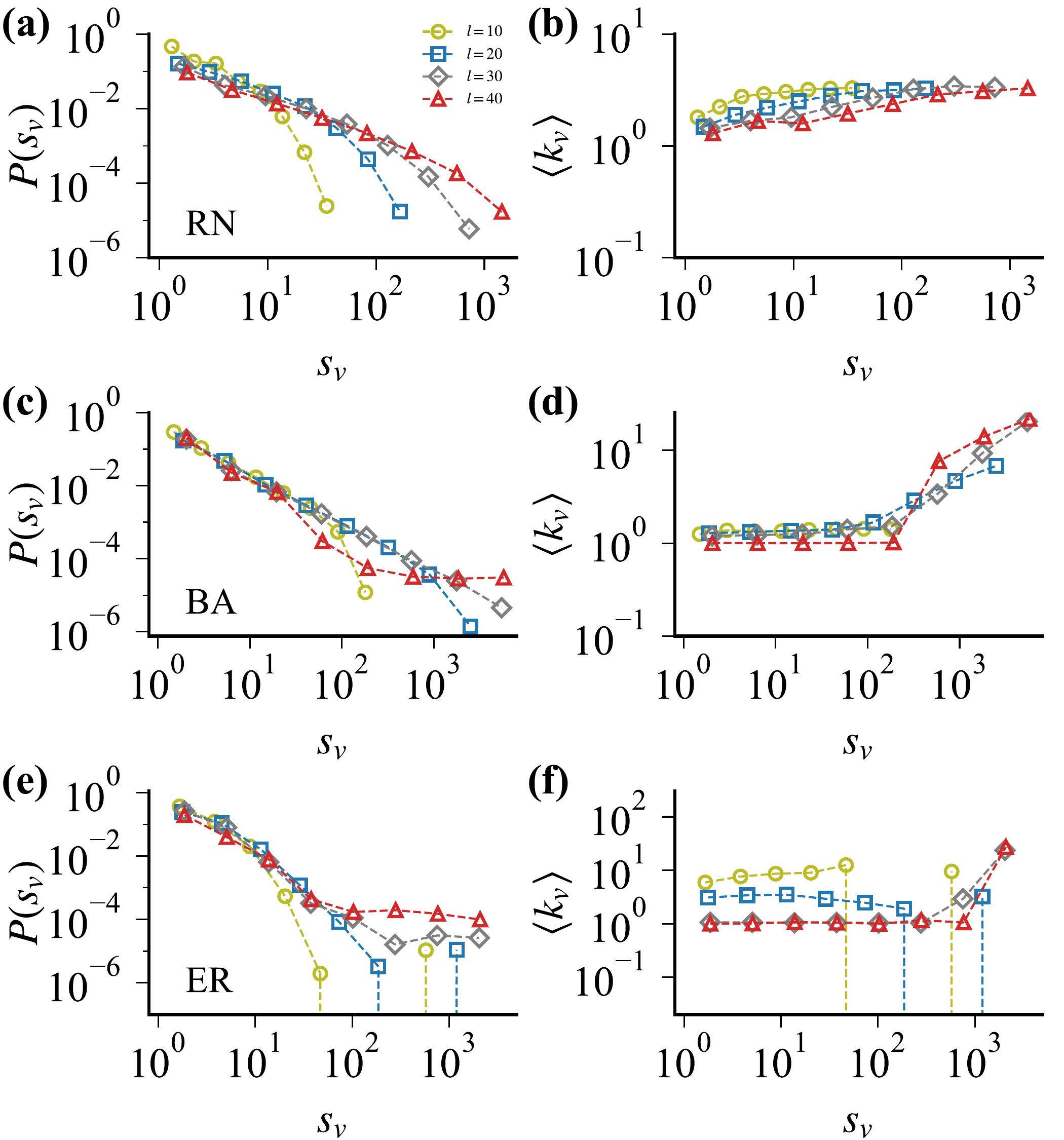} 
    \caption{Cluster-size and connectivity of supernodes under the binary LRG. Each row present the cluster-size distributions and the relation between degree and cluster-size during the binary LRG for (a,b) RN hypertrees, (c,d) BA hypergraphs, and (e,f) ER hypergraphs. (a,c,e) Distribution of the cluster-size of supervertex $s_v$, (b,d,f) degree versus cluster-size of supervertex.
     Each plot is obtained with $1000$ realization of size $V=E=2000$ through the binary LRG with $f=0.1$.
    }
    \label{fig:binary_clustersize}
\end{figure}

Under the binary LRG, the cluster-size distributions take power-law forms across RG steps while the slopes appear different between RN/BA hypertrees and ER hypergraphs [Fig.~\ref{fig:binary_clustersize}(a,c,e)]. Such broad cluster-size distributions for RN and BA hypertrees under the binary LRG are contrasted to the narrow ones under the EqLRG shown in Fig.~\ref{fig:clustersize}(a,c). The cluster-size distributions [Fig.~\ref{fig:binary_clustersize}(a,c,e)] take different forms from the degree distributions [Fig.~\ref{fig:binaryLRG}(d,h,l)].
The breakdown of the scaling between degree (also cardinality) and cluster size [Fig.~\ref{fig:binary_clustersize}(b,d,f)], attributable to the ad hoc formulation of the adjacency matrix under the binary LRG in Eq.~\eqref{eq:BI}, may explain the difference.

\subsection{EqLRG for pairwise networks in unipartite representation}
\label{sec:app_network}

The EqLRG can be applied also to pairwise networks in unipartite representation. Local-equilibrium node pairs $(i,j)$ are found by Eq. (9) and supernodes are formed as the connected components of those local-equilibrium pairs. The renormalized Laplacian and adjacency matrix are obtained by Eq. (5) and (7), treating $i$ and $j$ as general node indices without need to distinguish between vertices and hyperedges. 

\begin{figure}[h]
    \centering
    \includegraphics[width=0.8\textwidth]{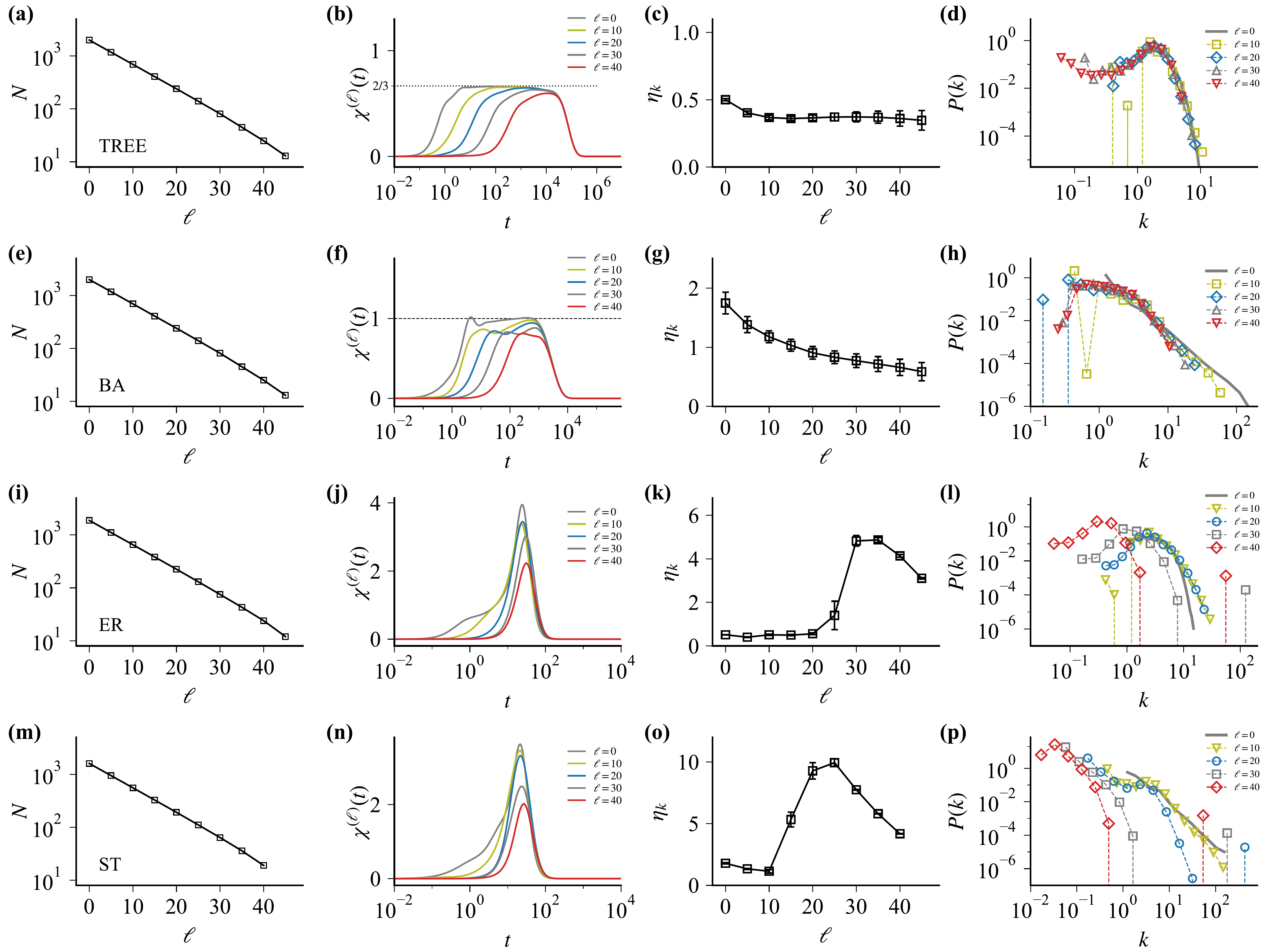} 
    \caption{Unipartite version of the EqLRG applied to (a-d) RN trees, (e-h) BA trees, (i-l) ER networks with $\bar{k}=3$, and (m-p) ST networks with $P(k)\sim k^{-2.5}$ and $\bar{k}=3$. The reduction fraction is $f=0.1$ and $\kappa=4$ are used for $N=2000$ vertices.}
    \label{fig:dyadic}
\end{figure}

Figure~\ref{fig:dyadic} shows the application results of this unipartite EqLRG to RN trees, BA trees, ST networks, and ER networks. These results are similar to those obtained for hypergraphs. For RN and BA trees, which initially have spectral dimensions $d_s=\frac{4}{3}$ and $2$ respectively, the plateau of the heat capacity remains at or shifts toward $\chi=\frac{2}{3}$ [Fig.~\ref{fig:dyadic}(b,f)]. Their degree heterogeneity indices also remain low or decrease, converging to similar values at late RG steps[Fig.~\ref{fig:dyadic}(c,g)], and their degree distributions remain close or evolve toward Poissonian forms[Fig.~\ref{fig:dyadic}(d,h)]. A notable difference in contrast to RN and BA hypertrees, however, is lack of an early-time peak at $\chi\approx \frac{1}{2}$ of the heat capacity, suggesting that this feature may be specific to the diffusion dynamics across vertices and hyperedges in the hypergraph setting adopted in this study.

For ER and ST networks, the heat capacity does not display a plateau [Fig.~\ref{fig:dyadic}(j,n)], indicating the absence of a well-defined spectral dimension for them. Their heterogeneity indices remain low over a range of RG steps until a rapid increase [Fig.~\ref{fig:dyadic}(k,o)] concomitant  with the appearance of humps in the degree distributions [Fig.~\ref{fig:dyadic}(l,p)] at late RG steps, which suggests the condensation of connections. Note that, unlike ER hypergraphs, ER networks do not exhibit clear power-law behavior in their degree distributions over a wide range of degree during the intermediate RG regime in accordance with the heterogeneity index remaining similar to the initial value throughout this regime. 

\subsection{Real-world hypergraphs}

\begin{figure}[h]
    \centering
    \includegraphics[width=\textwidth]{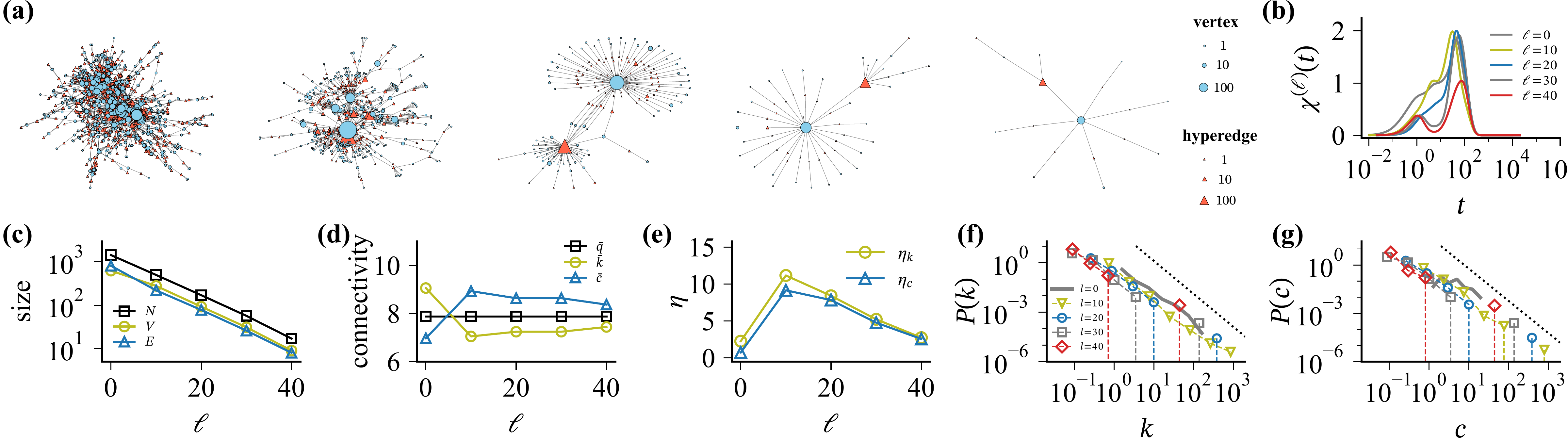} 
    \caption{Evolution of the drug-class hypergraph~\cite{BensenPaper,BensenData} under the EqLRG hypergraph. Vertices are class labels and hyperedges are drug products. A vertex-hyperedge connection means that the corresponding drug product belongs to the corresponding class.
    (a) Transformation of the hypergraph represented as a bipartite network from the original one ($\ell=0$) to the transformed ones at RG steps $\ell=10,20,30$ and $40$ (from left to right) with $f=0.1$.
    (b) Heat capacity $\chi(t)$ versus $t$ for different $\ell$’s.
    (c) Number of supernodes ($N$), supervertices ($V$), and superhyperedges ($E$) versus RG steps $\ell$.
    (d) Mean connectivity ($\bar{q}$), mean vertex degree ($\bar{k}$), and mean hyperedge cardinality ($\bar{c}$) versus $\ell$.
    (e) Heterogeneity indices $\eta_k$ and $\eta_c$ versus $\ell$.  
    (f) Distribution of renormalized supervertex degree $P(k)$ at different RG steps $\ell$.
    (g) Distribution of renormalized superhyperedge cardinaltiy $P(c)$. 
    }
    \label{fig:drug-class}
\end{figure}

\begin{figure}[h]
    \centering
    \includegraphics[width=\textwidth]{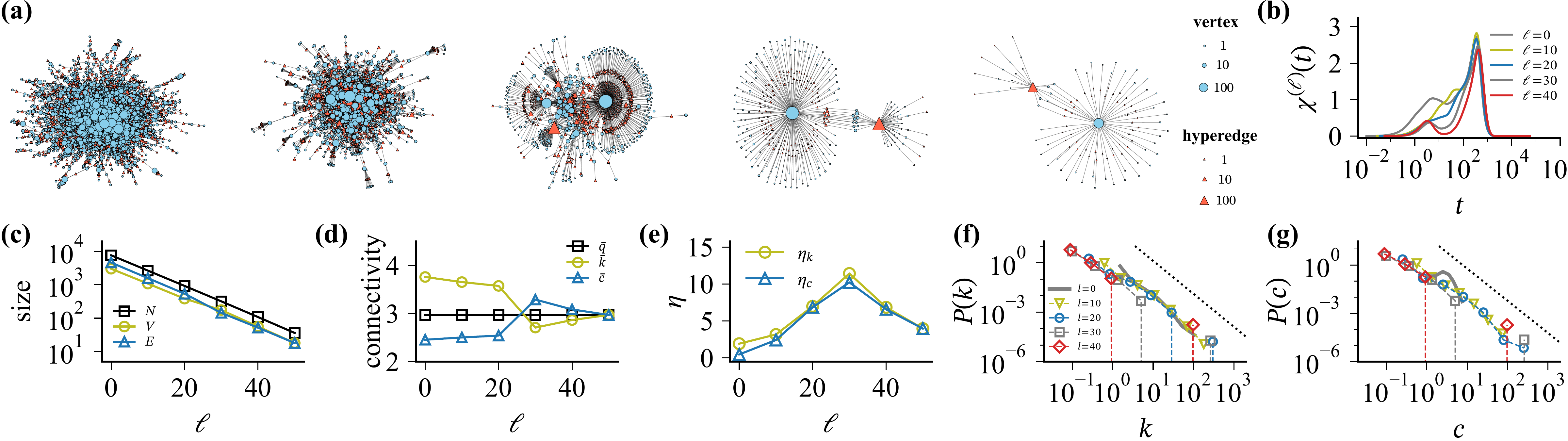} 
    \caption{Evolution of the coauthorship hypergraph from arXiv’s Quantitative Finance (q-fin) category from 2007
to 2016~\cite{ShapePaper,ShapeData} under the EqLRG. Vertices are researchers and hyperedges are preprints. Their connection means that a researcher is one of the authors of a preprint. (a-g) present the same plots as in Fig.~\ref{fig:drug-class}.
    }
    \label{fig:qfin}
\end{figure}

The EqLRG can also be applied to real-world hypergraphs. To showcase this applicability, in this section we present EqLRG results for two real-world hypergraphs. In Fig.~\ref{fig:drug-class}, we present the results for the drug-class hypergraph derived from the National Drug Code Directory and studied in Refs.~\cite{BensenPaper,BensenData}, where vertices are class labels and hyperedges represent drug products. In Fig.~\ref{fig:qfin}, shown are the results for the coauthor ship hypergraph derived from the arXiv Quantitative Finance (q-fin) category from 2007 to 2016~\cite{ShapePaper,ShapeData}, where vertices are authors and hyperedges are preprints. Both real-world hypergraphs have broad degree and cardinality distributions and exhibit behaviors similar to ST hypergraphs: they lack a well-defined spectral dimension, their heterogeneity indices initially increases with RG steps, and a ultra-hub vertex and hyperedge emerge at late stages of the RG process. 

\bibliography{references}